\documentclass[pra,aps,twocolumn,floatfix,showpacs]{revtex4-1}

\usepackage{graphicx,amssymb,amstext,amsmath,hyperref}

\usepackage[caption=false]{subfig}

\usepackage[usenames]{color}

\newcommand{\ket}[1]{\left| #1 \right>} 
\newcommand{\bra}[1]{\left< #1 \right|} 
\newcommand{\expectation}[1]{\left\langle #1 \right\rangle} 

\newcommand{\Tr}{\text{Tr}\,}

\begin{document}

\title{Thermodynamics of Quadrature Trajectories in Open Quantum Systems}
\author{James M. Hickey}
\author{Sam Genway}
\author{Igor Lesanovsky}
\author{Juan P. Garrahan}
\affiliation{School of Physics and Astronomy, University of Nottingham, Nottingham, NG7 2RD, United Kingdom}

\date{\today}

\begin{abstract}
We apply a large-deviation method to study the diffusive trajectories of the quadrature operators of light within a reservoir connected to dissipative quantum systems.  
We formulate the study of quadrature trajectories in terms of characteristic operators and show that in the long time limit the statistics of such trajectories obey a large-deviation principle.  We take our motivation from homodyne detection schemes which allow the statistics of quadrature operator of the light field to be measured.  We illustrate our approach with four examples of increasing complexity: a driven two-level system, a `blinking' three-level system, a pair of weakly-coupled two-level driven systems, and the micromaser.  We discuss how quadrature operators can serve as alternative order parameters for the classification of dynamical phases, which is particularly useful in cases where the statistics of quantum jumps cannot distinguish between such phases.   
The formalism we introduce also allows us to analyse the properties of the light emitted by quantum jump trajectories which fluctuate far from the typical dynamics.
\end{abstract}

\maketitle

\section{Introduction}
\label{sec:Intro}

This paper addresses questions about the nature of dynamical phase transitions and crossovers in open quantum systems from the perspective of observers able to make measurements on the environment.  We take our motivation from experiments using homodyne detection schemes~\cite{gerry2005introductory,Ariano2,gardiner2004} and study the time series or trajectories of quadratures of the light emitted from open quantum systems.   We go beyond recent work~\cite{Garrahan2010} which focused on trajectories of quantum jumps analyzed from the point of view of the counting statistics of, for example, photons entering the environment.  In this approach, identifying the average rate of photon emission as a dynamical order parameter allowed to uncover dynamical phase transitions and crossovers in a number of systems~\cite{Garrahan2011, CATES, Genway1, Budini2011, Garnerone2012, Li2011}.  A central idea in the works upon which we build was the introduction of a field conjugate to the number of emitted quanta, the so-called ``$s$-field''.  In equilibrium statistical mechanics, the observation of phase transitions is dependent on choice of the external field used to tune across a transition and the use of an appropriate order parameter;  in this work we demonstrate that different \emph{dynamical} order parameters and conjugate $s$-fields allow exploration of different \emph{dynamical} phase transitions.  Specifically we show that crossovers between dynamical phases may be observed from quadrature measurements which could not be found by counting photons, and vice versa.  We also extend our study to consider the effects of two $s$-fields, so that we can influence both light quadratures and quantum jumps.  This enables us to understand the relation between inferences from different dynamical observables; for example, we can find the typical quadrature measurements which would be made in rare quantum-jump trajectories, when either fewer or more photons are emitted than average.  

Our approach takes as a starting point the basic observation that (real-time) dynamics is more than statics. Sometimes dynamical behaviour, including transitions and crossovers between different dynamical regimes, can be understood from the properties of the stationary state.  But a more common occurrence is that dynamical fluctuations are only revealed via (often high-order) time-correlation functions, and not by static or one-time observables.  This suggests that for a proper statistical analysis of the dynamics of open systems a `statistical mechanics of trajectories' is needed.  Such an approach is the so-called $s$-ensemble method, which has proven useful for the study of classical many-body systems displaying complex cooperative dynamics such as glasses \cite{gale+:07,heja+:09,fredepl,speckpnas,toninelli}.  
The aim is to describe dynamical phases in terms of strictly dynamical order parameters and to classify these dynamical phases and the changes between them using a mathematical and conceptual framework analogous to that of equilibrium statistical mechanics.  This ensemble method for dynamics can therefore be thought of as a `thermodynamics of trajectories' approach \cite{Ruelle,Merolle,Lecomte,gale+:07,Garrahan2010}. 

Dynamical phase transitions are not limited to classical systems; transitions and crossovers have been discovered in a number of driven open quantum systems.  Famous examples include the laser~\cite{Scully1997}, whose behaviour close to threshold resembles a closed thermodynamic system in the neighbourhood of a continuous phase transition, and the micromaser~\cite{Walther2006,Haroche2006}, where a flux of atoms is coupled to an optical cavity mode and drives the cavity through a series of crossovers.  However more recently, dynamical phase transitions have also been discovered in a broad range of fields, spanning decohering spin channels~\cite{Alvarez2010}, information transport in complex systems~\cite{Aquino2011}, current fluctuations in isolated diffusive systems~\cite{Hurtado2011} and even the decoherence of tunnelling molecules~\cite{Coles2012}.   The $s$-ensemble method was recently applied to the study of ensembles of quantum-jump trajectories in open quantum systems~\cite{Garrahan2010}.   Although similar in spirit to ideas in full counting statistics~\cite{Lesovic1996, Nazarov2003, Mukamel, Flindt},  the $s$-ensemble approaches the problem from a different perspective and has now been developed for a variety of problems where dynamical phases are classified according to the counting statistics of quantum jumps \cite{Garrahan2010, Garnerone2012, Garrahan2011, Genway1, CATES, Budini2011, Li2011, madalin}.

\begin{figure}[!tb]
\includegraphics[scale=0.25]{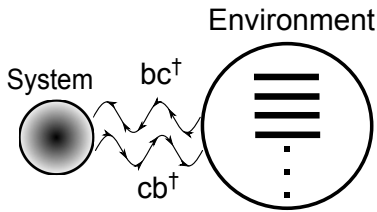}
\caption{A schematic diagram of the general Markovian system plus environment models we study.  The interaction between the system and the environment is of form $c{b}^{\dag}+b{c}^{\dag}$, where the operators $b^\dag$ ($b$) are creation (annihilation) operators of an harmonic oscillator bath and the operators $c$ and $c^\dag$ are system operators which are related to the Lindblad projectors of the system.}
\label{fig:gen}
\end{figure}

In this paper, we formulate the application of the $s$-ensemble to light quadratures by means of reduced characteristic operators.   Measuring quadratures of emitted light allows the quantum state of light emitted from an open system to be probed in a way not possible by counting photons.   For example, it provides an understanding of whether the light is in a coherent state, a squeezed state or another more exotic state.  We apply this method to a selection of open quantum systems, ranging from few-level quantum-optical systems to the micromaser, all of which have a stucture illustrated schematically in Fig.~\ref{fig:gen}.   In contrast to counting the number of emitted photons, as in studies of quantum-jump trajectories, we consider the temporal accumulation of quadratures of light emitted from a system into its environment.  We consider this a \emph{quadrature trajectory}.  Unlike the quantum-jump trajectories, the accumlation of light quadratures in the environment is a diffusive process.  However we can define a \emph{quadrature activity}, defined as a time-averaged light quadrature,  and use this as a dynamical order parameter to characterise dynamical phases in the space of quadrature trajectories.

\begin{figure}[!bt]$
\begin{array}{cc}
\includegraphics[scale=0.25]{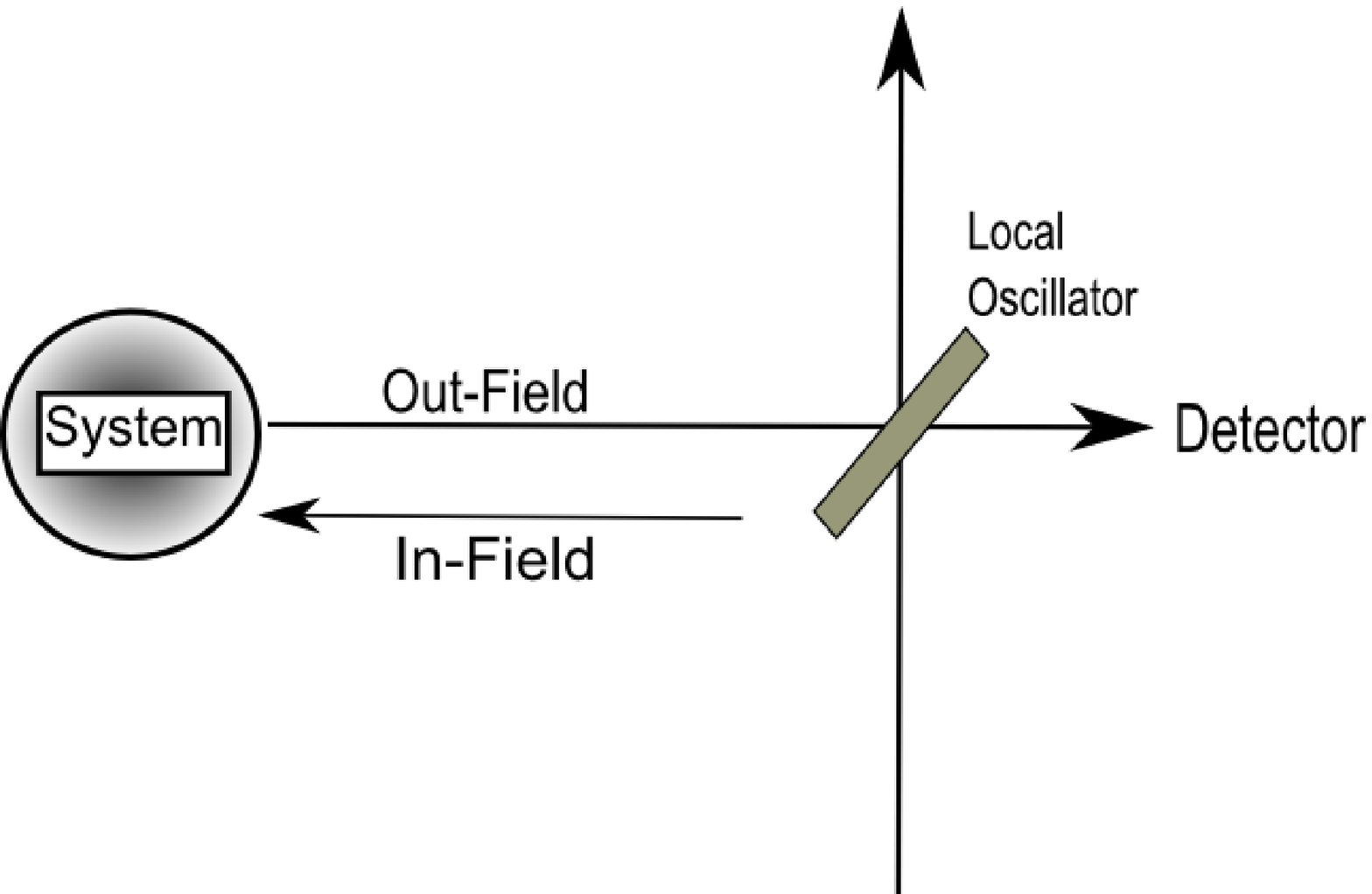}
\includegraphics[scale=0.125]{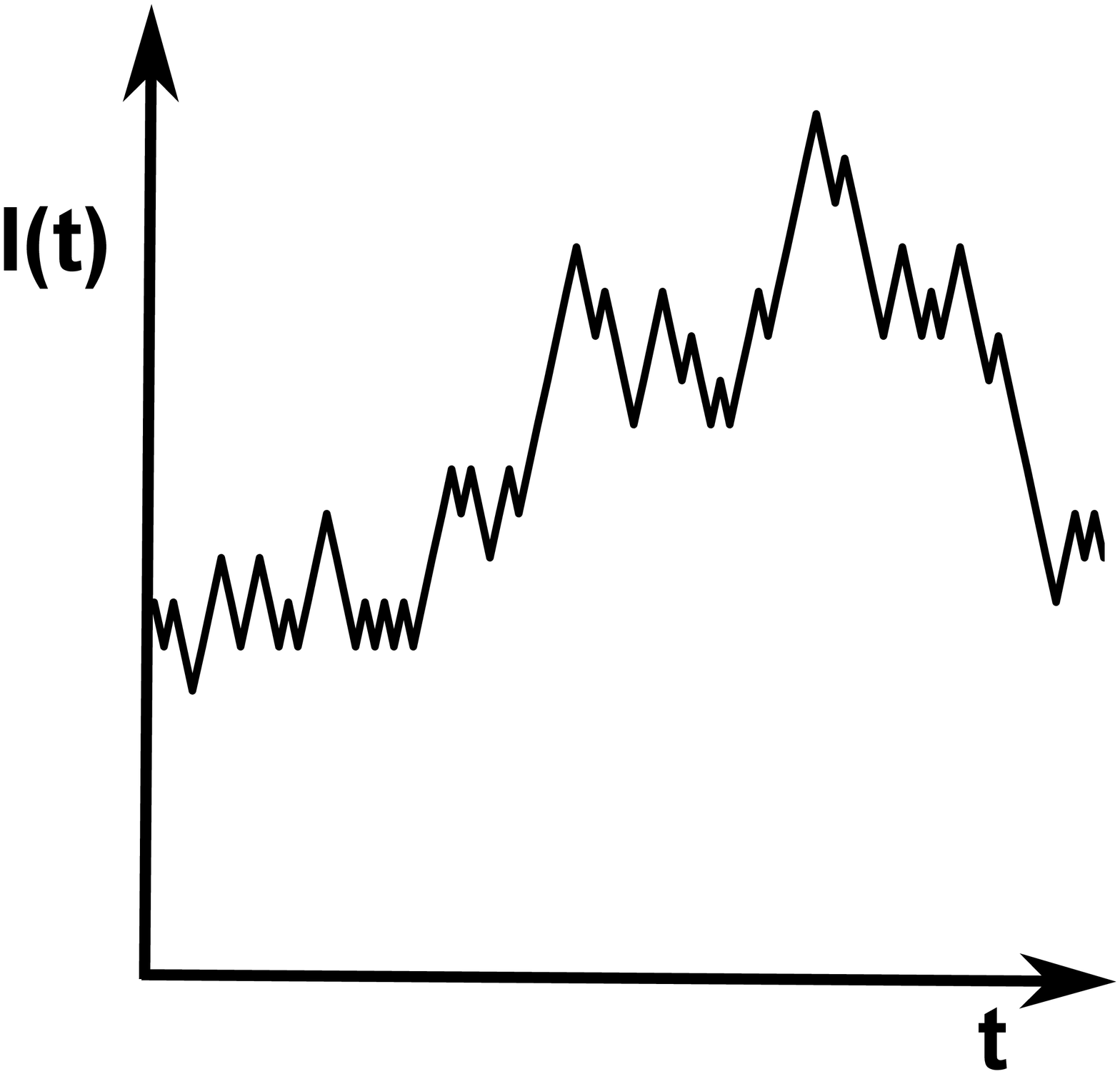}
\end{array}$
\caption{Shown (left) is a setup for homodyne detection and (right) a typical photocurrent, ${I}_{\text{homo}}$, plotted against time in such a measurement.    A simple homodyne scheme consists of an input beam and an output beam which is split between a detector and a strong coherent field which acts as a local oscillator.   If we denote ${b}_{\text{Out}}$ as the output lowering operator and $\mathcal{E}$ as the local oscillator strength, the total transmitted field is ${b}^{h}_{\text{Out}}=\mathcal{E}+{b}_{\text{Out}}$.  Looking at the photocurrent ${I}_{\text{p}}=d{b}^{\dag h}_{\text{Out}}{b}^{h}_{\text{Out}}/dt$ one may define the homodyne current ${I}_{\text{homo}}= \frac{{I}_{p}}{\mathcal{E}}-\mathcal{E}$ in the limit of infinite local oscillator strength. By expanding ${I}_{p}$ one finds that the homodyne current is a direct measure of the quantity $b+b^\dag$, which is directly related to the $X$-quadrature.   Ref.~\cite{Ariano2} contains further details about the various homodyne detection schemes which allow $X$-quadrature measurements.   In this work. we use the input and output fields to form the reservoir we depict in Fig.~\ref{fig:gen}.  }\label{fig:3}
\end{figure}

Our motivation for studying the statistics of quadratures is inspired by the experimental technique of homodyne detection.   The $X$-quadrature trajectories of emitted light are directly related to the \emph{homodyne current}~\cite{gerry2005introductory,Ariano2,gardiner2004},  as described in Fig.~\ref{fig:3}.  (It is also sensible for us to study other quadratures, since these are accessible via a change in the driving Hamiltonian, which we will discuss later in Sec.~\ref{sec:level2}.)  The statistics of quadrature trajectories provide a more natural probe of the system dynamics than quantum-jump trajectories and we will show that such measurement schemes allow exploration of dynamical phases in a variety of systems.

Beyond examining dynamical phases identified by quadrature activity, we construct marginal probability distributions for general quadrature operators at all angles in phase space.  Using these marginals, we reconstruct Wigner distributions~\cite{gardiner2004} via the inverse radon transform, to find the state of the emitted light.  We extend our studies to examine the typical quadrature trajectories of systems  biased by the number of photon emissions.  This allows us to understand the nature of light emitted from a system for all quantum-jump trajectories, whether \emph{rare} or \emph{typical} with respect to the number of emitted photons.   More generally we study the trajectories of a particular observable after having biased the system towards rare trajectories of another (generally non-commuting) observable.  This technique also allows us to identify the appropriate dynamical order parameters to understand the dynamical phases in different open systems.

The paper is structured as follows.  In the following section we introduce the formalism of the $s$-ensemble and describe the It\={o} calculus methods which we employ for the stochastic description of the environment we use in this paper.  We further develop generalised master equations for the dynamics of quadrature-biased systems and systems biased towards rare trajectories of both quantum-jumps and quadratures.  We also discuss use of the Wigner function to analyse our results.  In subsequent sections we present our results.  In Sec.~\ref{sec:level2} we discuss a simple driven two-level system and in Sec.~\ref{sec:level3} we extend our analysis to a driven three-level system.  In Sec.~\ref{sec:2L2C} we study a pair of coupled two-level systems and demonstrate that measuring quadrature trajectories allows identification of dynamical phases which are hidden in photon-counting experiments.  Finally, in Sec.~\ref{sec:MICRO} we give a concise account of quadrature trajectories in the micromaser, providing insights via a mean-field theory as well as exact numerical diagonalisation.  In Sec.~\ref{sec:Conc} we give our conclusions.

\section{Formalism and Theoretical Background}

Several recent studies have applied the $s$-ensemble method to study the statistics of quantum trajectories in open quantum systems using a thermodynamic description~\cite{Touchette2009,Lecomte:2007fk}.  In this section we will give an account of the $s$-ensemble formalism for quantum trajectories.  We discuss the specific examples of counting photons and measuring the quadratures of light entering the environment around an open quantum system.    First, in Section~\ref{sec:sensemble} we discuss the general application of the $s$-ensemble in the study of the statistics of an observable $Q$. We then reformulate these principles so their application to open systems is clear. Section~\ref{sec:inout} discusses the open quantum system methods which we use, which are applied to the $s$-ensemble in Sections~\ref{sec:s1} and~\ref{sec:s2}. 

\subsection{An $s$-ensemble for open quantum systems}
\label{sec:sensemble}

We begin by considering a particular measurable quantity $Q$ associated with a quantum trajectory of a quantum system coupled to a reservoir.  (For example, $Q$ could be the number of emitted photons in a time $t$, or other quantities which we define below.)  Projecting the system density matrix $\rho(t)$ on to the subspace where $Q$ takes a particular value, one may define a reduced density matrix ${\rho}^{(Q)}_{t}$.  
The probability of such a realisation occuring in time $t$ is then given by $P_{t}(Q) = \Tr[{\rho}^{(Q)}(t) ]$ and, after long times when the system reaches a steady state, this takes a large-deviation (LD) form:
\vspace*{-5mm}
\begin{center}
\begin{equation}\label{eq:8e}
P_{t}(Q) \simeq {e}^{-t\phi(Q/t)}
\end{equation}
\end{center}
where $\phi(Q/t)$ contains all the information about the probability distribution of $Q$ at long times. 
Alternatively the statistics may be described introducing a moment generating function associated with these probabilities.  This method proceeds by introducing another density matrix ${\rho}^{s}(t)$ defined by the Laplace transform
\vspace*{-5mm}
\begin{center}
\begin{equation}
\label{eq:rhos}
{\rho}^{s}(t) = \int\rho^{\left(Q\right)}(t){e}^{-sQ}dQ
\end{equation}
\end{center}
from which we define the moment generating function
\begin{equation}
\label{eq:Z}
Z_t(s) = \Tr \rho^s(t) = \int P_t(Q) \,e^{-sQ} dQ\,.
\end{equation}
In both~\eqref{eq:rhos} and~\eqref{eq:Z}, the integrals should be replaced with sums if $Q$ is a discrete quantity, as is the case when counting photons.  A LD form 
\begin{equation}
Z_t(s)\simeq e^{t\theta(s)}
\end{equation}
is found at long times, with LD functions $\theta_Q(s)$ and $\phi(Q/t)$ related by Legendre transform, $\phi(Q/t) = -\text{min}_{\text{s}}(\theta_Q(s)+(Q/t)s)$.  The full statistics of $Q$ are therefore contained within $\theta_Q(s)$.
The density matrix $\rho^s(t)$ describes, when $s\ne 0$, rare trajectories where $Q$ is far from the mean.  In counting processes, where $Q$ is bounded from below by zero, the rare trajectories can be separated into \emph{more active} trajectories when $s<0$ and \emph{less active} trajectories when $s>0$.  These correspond respectively to trajectories with a larger or a smaller number of counts than in the $s=0$ physical dynamics.

There are two immediate advantages to this $s$-ensemble approach.  Firstly, as the LD function $\theta_Q(s)$ is the largest real eigenvalue of a superoperator which generates the $s$-biased dynamics of $\rho^s(t)$ (which we derive in the following sections), evaluation of the statistics of trajectories can be straightforward.  Of particular interest are the first and second moments, $q_s$ and $\Delta q_s^2$, which may be found directly from the LD function via
\begin{eqnarray}
q_s &\equiv & \frac{\expectation{Q}_s}{t}  =  -\frac{\partial\theta_Q}{\partial s}(s) \label{eq:qs}\\ 
\Delta q_s^2 &\equiv & \frac{\expectation{Q^2}_s - \expectation{Q}_s^2}{t}  =  \frac{\partial^2\theta_Q}{\partial s^2}(s)\,,\label{eq:qs2}
\end{eqnarray}
where the $s$ subscripts indicated that the expectation values are taken with respect to the ensemble of trajectories biased by $e^{-sQ}$.  Secondly, this method provides a thermodynamic formalism for non-equilibrium processes.  The LD functions $\theta_Q(s)$ and $\phi(Q/t)$ are analogues of free energy and entropy densities, with $s$ the conjugate \emph{intensive} field to the \emph{time-extensive} $Q$.  Furthermore, the so-called activity $q_s$ may be used as a dynamical order parameter to distinguish dynamical phases, whose boundaries may be crossed by tuning the parameter $s$, or other system parameters.  Indeed, such phase boundaries are crossed when an $s$-derivative of $\theta_Q(s)$ becomes discontinuous. 

So far, the $s$-ensemble studies of open quantum systems have explored the thermodynamics of quantum-jump trajectories associated with counting quanta emitted from a system into a Markovian environment.  In these cases the quantity $Q$ is, for example, the number of photons emitted (\emph{i.e.} the number of quantum jumps), $K$, from a system in unit time.  Defining reservoir ladder operators $b$ and $b^\dag$ in the Heisenberg representation, the quantity $K$ corresponds integrating over time the observable $b^\dag b$ .  Biasing these trajectories, we obtain a LD function $\theta_K(s)$ from which we extract the dynamical order parameter $k_s = \expectation{K}_s/t$. 

\begin{figure}[hb]
\includegraphics[scale = 0.25]{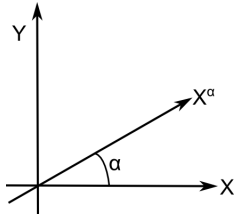}
\caption{The quadrature operators define axes of an optical phase space. The generic quadrature operator ${X}^{\alpha}$ may be viewed as a rotation of the X-quadrature axis as shown above.}
\label{fig:2}
\end{figure}

In this paper, we develop the $s$-ensemble further to study the statistics of the quadratures of light entering a Markovian bath from various quantum systems.  Choosing $Q$ to be the $X$- and $Y$-quadratures of the light, corresponding to bath operators $(b+b^\dag)/2$ and $i(b-b^\dag)/2$, we will use corresponding \emph{quadrature activities} $x_s = \langle X\rangle_s/t$ and $y_s=\langle Y\rangle_s/t$ when applying $s$-fields to bias the dynamics towards rare quadrature trajectories.  In much the same way as the photon activity $k_s$ can be associated with the average number of photons emitted in a time $t$, the quadrature activities $x_s$ and $y_s$ are time averages of the $X$- and $Y$-quadratures of the light leaving the system.  We will gain yet further insight into the dynamics of the systems we study by examining general quadratures of the form 
\begin{equation}
\label{eq:generalquad}
X^\alpha = \cos{\alpha} ~ X + \sin{\alpha} ~ Y
\end{equation}
where the angle $\alpha$ in phase space is illustrated in Fig.~\ref{fig:2}.
For these general quadratures, we can consider marginal distributions $P_t(X^\alpha)$ associated with general quadratures, which can be extracted from the relevant LD functions $\theta_{X^\alpha}(s)$ and their $s$-derivatives using Eqs.~\eqref{eq:qs} and~\eqref{eq:qs2}.  

In the following subsections we will introduce the mathematical formalism which allows us to construct $s$-biased ensembles and show how the appropriate LD functions are derived.
We will also explore doubly-biased ensembles where we study the typical trajectories of one observable for ensembles of trajectories biased by another observable.   We develop this formulation in Section~\ref{sec:s2}, but we first discuss how the LD functions for (singly-) biased ensembles are derived. 
 
\subsection{Generalized Master equations, It\={o} Calculus and Quadratures}
\label{sec:inout}
Previous studies on quantum-jump trajectories~\cite{Garrahan2010,Garrahan2011} have demonstrated that one may identify the LD function $\theta_Q(s)$ as the largest eigenvalue of an $s$-modified master equation obeyed by ${\rho}^{s}$,
\vspace*{-5mm}
\begin{center}
\begin{equation}
\label{eq:Ws1}
\dot{\rho}^{s} = {\mathcal{W}}_{s}({\rho}^{s}),
\end{equation}
\end{center}
where ${\mathcal{W}}_{s}$ is the generator for the $s$-biased dynamics. We will outline how to construct formally these generalized master equations in terms of an It\={o} calculus and characteristic operators, before proceeding to apply them to the study of quadratures in Sec.~\ref{sec:s1}.

We consider the system to be weakly coupled to a reservoir in the Markovian regime~\cite{gardiner2004}; within this picture the reservoir is treated as a bath of harmonic oscillators which effectively act as a white noise source to our system. This type of reservoir admits a stochastic description whereby the effects of reservoir ladder operators $b(t )$ and ${b}^{\dag} (t )$, with commutator $[b (t ),{b}^{\dag}  (t' )] = \delta (t-t')$ are described by It\={o} increments which obey a quantum It\={o} calculus~\cite{gardiner2004}.  We next sketch our method within this formalism; we give full details of the derivation in Appendix~\ref{sec:Der1}. 

For the rest of the paper we consider the reservoir to be an unsqueezed vacuum \cite{Note1}. We introduce temporal increments ${dB }^{\dag}\left( t \right)$ and $dB \left( t \right )$ which are defined in terms of the reservoir ladder operators $b(t)$ and $b^\dag (t)$ by
\vspace*{-5mm}
\begin{center}
\begin{align}\label{eq:1e}
&{dB}^{\dag}(t)= \int^{t+dt}{{b}^{\dag}(t')\, dt'}-\int^{t}{{b}^{\dag}(t')\, dt'},\nonumber\\
&dB(t)=\int^{t+dt}{b(t')\, dt'}-\int^{t}{b(t')\, dt'}.
\end{align}
\end{center}
and are evaluated in the immediate future. The dynamics are found using a stochastic density matrix $R_(t)$ whose It\={o} increment is related to the system Hamiltonian, $H$, the reservoir increments defined above and the Lindblad operators of the system $\left({L}_{i} \right)$:
\vspace{-5mm}
\begin{center}
\begin{align}\label{eq:2e}
&d\rho\left(t \right)= -i[H,\rho]dt-\frac{1}{2}\sum_{i}\{{L}^{\dag}_{i}{L}_{i},\rho\}dt\\ 
&+\sum_{i}{dB}^{\dag}\left(t \right){L}_{i}\rho\left(t \right){L}^{\dag}_{i}dB\left(t \right)\nonumber\\
&+\sum_{i}{dB}^{\dag}\left(t \right){L}_{i}\rho(t)\nonumber \\
&+\sum_{i}\rho\left(t \right)dB\left(t \right){L}^{\dag}_{i}.\nonumber
\end{align}
\end{center}
Tracing out the reservoir degrees of freedom, one obtains the familiar Lindblad master equation~\cite{gardiner2004,gorini1976,Lindblad} for the (trace-preserving) system dynamics:
\vspace*{-5mm}
\begin{center}
\begin{eqnarray}
\dot{\rho^0}(t)&=&\mathcal{L}\left(\rho^0\right)\nonumber\\
												&=&-i[H,\rho_0]-\sum_{i}\{{L}^{\dag}_{i}{L}_{i},\rho^0 \}+ \sum_{i}{L}_{i}\rho^0{L}^{\dag}_{i}\label{eq:3e}
\end{eqnarray}
\end{center}
where $\{\bullet,\bullet\}$ is the anti-commutator and $\rho^0=\Tr_\text{Res}(\rho)$ is the density matrix of the system, with $\text{Tr}_{\text{Res}}$ denoting the trace over the reservoir Hilbert space.

To modify this trace-preserving scheme to the desired $s$-biased generalized master equation for trajectories of $Q$, we introduce a characteristic operator, ${V}^{s}_{Q}(t )$.  The corresponding stochastic increment d$Q$ is related to the reservoir increments in Eq.~\eqref{eq:1e}.  The characteristic operator is defined by
\vspace*{-5mm}
\begin{center}
\begin{equation}\label{eq:charQ}
{V}^{s}_{Q}\left(t\right) = \text{exp}\left(-s\int^{t} dQ({t}')\right).
\end{equation}
\end{center}
From this operator we identify $s$-biased reduced density matrix as
\vspace*{-5mm}
\begin{center}
\begin{equation}\label{eq:rhoQ}
{\rho}^{s} = \text{Tr}_{\text{Res}}\left({V}^{s}_{Q}\left(t\right)\rho(t)\right)\,.
\end{equation}
\end{center}
Using Eq.~\eqref{eq:rhoQ} one may derive Eq.~\eqref{eq:Ws1} by finding the increment of ${\rho}^{s}$ in terms of the increments of ${V}^{s}_{Q}$ and $\rho$.  Then, application of the It\={o} calculus of d$Q$ in terms of reservoir increments~\eqref{eq:1e} allows the trace to be taken over the reservoir degrees of freedom (see Appendix~\ref{sec:Der1}). This method of deriving ${\mathcal{W}}_{s}$ relies upon the existence of a complementary stochastic reservoir operator to the  system operator of interest.  We discuss these operators next for the cases of quantum jumps and quadratures.

We turn our attention to statistics of quantum-jump trajectories, before moving on to quadrature trajectories, where $Q=K$.  A quantum-jump trajectory is the time record of the number of jump events $K$ over a time $t$.  We formulate the stochastic process associated with jump trajectories in terms of an increment $dK$. In terms of the reservoir increments this jump increment is defined as $dK = d{B}^{\dag}dB/dt$ which has eigenvalues of 0 and 1 within the interval $[t,t+dt]$.  Although $dK$ tells us how many photons were emitted into the bath in the interval $[t,t+dt]$, it tells us nothing else about the form of the light emitted.  We address questions about the quantum state of emitted light by looking at the statistics of the quadrature trajectories of the light emitted into the reservoir. We use the $X$- and $Y$-quadratures to define coordinate axes of the optical phase space (see Fig.~\ref{fig:2}).  We will also study the general quadrature operator where the observable is $Q={X}^{\alpha}_{t}$, defined in Eq.~\eqref{eq:generalquad}, where $\alpha$ is the polar angle.  The stochastic increments associated with these operators, $dX^\alpha(t)$, are defined by
\vspace*{-5mm}
\begin{center}
\begin{equation}\label{eq:6e}
d{X}^{\alpha}(t)=\frac{1}{2} \left({e}^{-i\alpha}d{B}(t) +{e}^{i\alpha}d{B}^{\dag}(t) \right).
\end{equation}
\end{center}

Evaluation at $\alpha = 0$ and $\pi/2$ yields the increment for the familiar $X$- and $Y$-quadratures. Although $dX^\alpha(t)$ takes the same form for all $\alpha$, the form of the stochastic process $dK$ associated with the jump trajectories is very different. Therefore, for the rest of this paper we use notation where $K$ is biased by a conjugate field $s'$, while the quadratures $X^\alpha$ are biased by $s$.  With this notation, we define characteristic operators of the form~\eqref{eq:charQ} associated with each ensemble of trajectories,
\vspace*{-5mm}
\begin{center}
\begin{align}\label{eq:14e}
& {V}^{s}_{X^\alpha}(t) = \text{exp}\left(-s\int^{t} d{X}^{\alpha}({t}')\right) \nonumber \\
& {V}^{{s}'}_{K}(t) = \text{exp}\left(-s\int^{t} dK({t}')\right).
\end{align}
\end{center}
With these definitions we will construct generalized master equations for each ensemble of trajectories in Sec.~\ref{sec:s1}, and we will proceed further in Sec.~\ref{sec:s2} to ask about the typical statistics of one process in the biased ensemble of the other.

\subsection{An $s$-Ensemble for Quadrature Trajectories}
\label{sec:s1}

We now turn to the statistics of the trajectories associated with the quadrature operators.  We consider projections of the density matrix $\rho(t)$ on to the subspace where $X^{\alpha}$ takes a specific realisation.  By this we mean that the time-integrated quadrature for light entering the bath up to time $t$ takes a specific value $X^\alpha$.  We denote a corresponding projected reduced density matrix by ${\rho}^{\left({X}^{\alpha}\right)}_{t}$.  As described in Section~\ref{sec:sensemble}, this is related to an $s$-biased reduced density matrix ${\rho}^{s}(t)$, defined in Eq.~\eqref{eq:rhos} with $Q=X^\alpha$, via the Laplace transform
\vspace*{-5mm}
\begin{center}
\begin{equation}\label{eq:7e}
{\rho}^{s}(t) = \int\rho^{\left({X}^{\alpha}\right)}(t){e}^{-s{X}^{\alpha}}d{X}^{\alpha}.
\end{equation}
\end{center}
which upon taking the trace gives us the moment generating function associated with these diffusive probabilities $Z_{t}(s) = \Tr \rho_s(t) \simeq {e}^{t\theta_{X^\alpha}(s)}$, where the LD form is valid at long times. The \emph{activity} associated with the quadratures is ${x}^{\alpha} = \langle {X}^{\alpha} \rangle/t$.  This, and the second moment $\Delta x^\alpha {}^2$ are found from the derivatives of the LD function $\theta_{X^\alpha}(s)$ defined in Eqs.~\eqref{eq:qs} and~\eqref{eq:qs2} with $Q=X^\alpha$.  
 This LD function is identified as the largest real eigenvalue~\cite{Touchette2009,Eckmann} of the generalized master equation
\vspace*{-5mm}
\begin{center}
\begin{align}\label{eq:10e}
&{\dot{\rho}}^{s}(t)={\mathcal{W}}_{s}\left({\rho}^{s}\right) \nonumber \\ &=\mathcal{L}\left({\rho}^{s}\right)-\sum_{i}\frac{s}{2}\left({e}^{-i\alpha}{L}_{i}{\rho}^{s}+{e}^{i\alpha}{\rho}^{s}{L}^{\dag}_{i}\right)+\frac{{s}^{2}}{8}{\rho}^{s}\,.
\end{align}
\end{center} 
We derive this generalised master equation using Eqs.~\eqref{eq:2e} and~\eqref{eq:charQ}, where tracing out the environment is performed within the It\={o} calculus formalism, the details of which are presented in Appendix~\ref{sec:Der1}.   When $s \to 0$, the superoperator collapses to the trace-preserving Liouvillian $\mathcal{L}$, Eq.~\eqref{eq:3e}.  Away from $s=0$ the $s$-field biases the dynamics towards rare trajectories of the system.  The properties of $s$-modified master equations are discussed in Ref.~\cite{Garrahan2010,Garrahan2011}.  
Before presenting our results for various model systems, we discuss doubly-biased ensembles.

\subsection{Quadratures of Quantum Jump Trajectories: Doubly-Biased Ensembles}
\label{sec:s2}

Thus far we have provided the theoretical formalism for an $s$-ensemble of quadrature trajectories.  Before presenting results for a variety of systems, we introduce a further interesting problem in this section concerning the trajectories of already-biased ensembles.  Consider a system biased towards rare trajectories with, for example, more quantum jumps than the $s'=0$ average.  We now ask what are the quadratures for these trajectories which are, for example, \emph{more active} with respect to the number of emitted photons? 
In order to answer this question we need to introduce two counting fields, 
$s'$ conjugate to the number of emitted photons, $K$, and $s$ conjugate to quadratures $X^\alpha$.  

The derivation of doubly-biased ensembles is an extension of the derivation in Sec.~\ref{sec:s1}, using Eq.~\eqref{eq:2e} with characteristic operators and tracing out the reservoir.   The characteristic operators associated with the quadrature increment and photon increment are respectively ${V}^{s}_{X^\alpha}\left(t\right)$ and ${V}^{s'}_{K}\left(t\right)$.  With these we can define a new $s$ and $s'$ biased density operator, $\rho^{ss'}$, which incorporates information on the statistics of both ensembles of trajectories,
\vspace*{-5mm}
\begin{center}
\begin{equation}\label{eq:15e}
{\rho}^{ss'}(t)={\text{Tr}}_{\text{res}}\left({{V}^{s'}_{K}}^{\frac{1}{2}}{V}^{s}_{X^\alpha}{{V}^{s'}_{K}}^{\frac{1}{2}}\rho(t) \right).
\end{equation}
\end{center}
The ordering of the characteristic operators is important due to the non-commutability of the two observables.  As defined, Eq.~\eqref{eq:15e} is interpreted as biasing the jump trajectories and within these $s'$-biased ensembles measuring the quadrature realisations.  We show in Appendix~\ref{sec:Der1} how It\={o} calculus is used to calculate the It\={o} increment of this new double-biased density operator.  (We formally trace out the reservoir using the It\={o} tables in Eqs.~\eqref{eq:13e} and~\eqref{eq:4e}.)  We find this new doubly-biased master density operator obeys a new generalised master equation which reads 
\vspace*{-5mm}
\begin{center}
\begin{eqnarray}\label{eq:16e}
{\rho}^{ss'}(t)&=& \mathcal{L}\left({\rho}^{ss'}\right) +\left({e}^{-s'}-1\right)\kappa c{\rho}^{ss'}{c}^{\dag}\\ 
&&+\frac{{s}^{2}}{8}{\rho}^{ss'}-\frac{s\sqrt{\kappa}{e}^{-\frac{s'}{2}}}{2}\left({e}^{-i\alpha}c{\rho}^{ss'}+{e}^{i\alpha}{\rho}^{ss'}{c}^{\dag}\right)\,. \nonumber
\end{eqnarray}
\end{center}
Notice that if $s$, $s'\to 0$ then the left-hand side reduces to the (trace-preserving) Liouvillian. Once again we may encapsulate the total long time trajectory statistics using the largest eigenvalue, ${\theta}_{K,X^\alpha}(s,s')$, of the generator of these statistics.  Furthermore we note that ${\theta}_{K,X^\alpha}(s,s')$ necessarily reduces to the LD function for $s'$-biased jump trajectories, ${\theta}_{K}(s')$, in the limit $s\to 0$~\cite{Garrahan2010,Genway1}. To examine the typical quadratures of quantum-jump biased systems we examine the derivatives of ${\theta}_{K,X^\alpha}(s,{s'})$, with respect to $s$, evaluated in the limit $s\to 0$.  Note that similar doubly-biased ensembles may be created by choosing different characteristic operators in Eq.~\eqref{eq:15e}.  For example, switching the characteristic operators in Eq.~\eqref{eq:15e} allows us to find the statistics of quantum jumps in quadrature-biased ensembles.

Next, we put the formalism into practice in the following sections with studies of few-level quantum-optical systems and a system with more complex dynamics, the micromaser.


\section{Driven two-level open systems}
\label{sec:level2}

\begin{figure}[h]
\includegraphics[height=2cm]{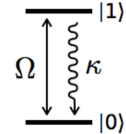}
\caption{Schematic diagram of a laser-driven two-level system coupled to a vacuum reservoir. }
\label{fig:1}
\end{figure}

In this section we present our results for a simple quantum-optical system consisting of a dissipative two-level system driven by a laser, shown schematically in Fig.~\ref{fig:1}.  We begin by examining the $X$- and $Y$-quadrature statistics of the system and compare it with $s$-ensemble studies for the jump trajectories studied in~\cite{Garrahan2010}. After examining both analytic and numerical forms of the LD function we turn our attention to the doubly-biased statistics of the system.  We finish by constructing plots of time-independent probability distributions, at various $s'$, associated with the $X^\alpha$ which we refer to as phase portraits before examining the Wigner functions of this system. 

For the driven two-level system, the generalized master operator, ${\mathcal{W}}_{s}$, is of the form in Eq.~\eqref{eq:10e}, where now we have one pair of Lindblad operators $L$ and $L^\dag$, where $L=\sqrt{\kappa}c$, with $\kappa$  the decay rate and $c$ the lowering operator $\ket{0}\bra{1}$. We drive the system with a laser polarised in the $x$-direction, which introduces a driving term $\propto \sigma_{x}$ in the two-level system Hamiltonian,
\vspace*{-5mm}
\begin{center}
\begin{equation}
H=\Omega\left(c + {c}^{\dag}\right),
\end{equation}
\end{center} 
where $\Omega$ is the Rabi frequency. Here we consider the specific choice $\kappa=4\Omega$ as this has been shown to be an interesting parameter choice in previous works~\cite{Garrahan2010}, and for further reasons we discuss below.  We study the $X$- and $Y$-quadratures specifically.   In this regime although the LD function does not take on a straightforward form, writing ${\mathcal{W}}_{s}$ in matrix form we numerically diagonalise it and identify its largest real eigenvalue as the LD function. Although no concise analytical form exists for the LD function, we may Taylor expand the LD functions about $s=0$ to see how the statistics of the physical system behave.  For $X$- and $Y$-quadratures we find the expansions
\vspace*{-5mm}
\begin{center}
\begin{align}\label{eq:17e}
&{\theta}_{X}\left(s\right)= \frac{7{s}^{2}}{24} -\frac{{s}^{4}}{36\Omega}+\frac{5{s}^{6}}{648{\Omega}^{2}}-\frac{{s}^{8}}{348{\Omega}^{3}}+\mathcal{O}\left({s}^{9}\right) \nonumber \\
&{\theta}_{Y}\left(s\right)= \frac{2s\sqrt{\Omega}}{3} +\frac{{s}^{2}}{8}-\frac{2{s}^{3}}{81\sqrt{\Omega}}+\frac{2{s}^{4}}{243\Omega}\nonumber\\ &\quad-\frac{8{s}^{6}}{6561{\Omega}^{2}}+\frac{10{s}^{7}}{19683{\Omega}^{5/2}}+\mathcal{O}\left({s}^{9}\right).
\end{align}
\end{center}

\begin{figure}[hbt]
\includegraphics[scale=1]{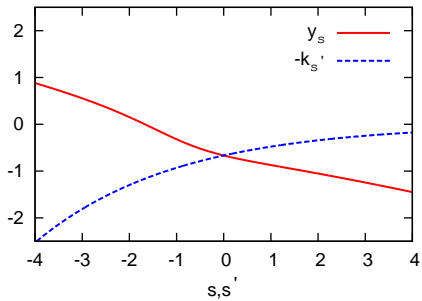}
\caption{A plot of the $Y$-quadrature activity $|y_s|$ and minus the quantum-jump activity $-k_{s'}$ for the two-level system when biased by the appropriate fields $s$ and $s'$.  Note that the order parameters coincide for the physical dynamics at $s=s'=0$. }
\label{yandk}
\end{figure}

Firstly we examine the form of the LD functions close to $s=0$. From the Taylor expansion~\eqref{eq:17e}, we see that that ${\theta}_{X}\left(s\right)$  to $\mathcal{O}\left({s}^{9}\right)$, is symmetric about $s=0$.  Immediately we find that the $X$-quadrature activity for physical $s=0$ dynamics is $x_{s=0}=0$.
This symmetry is not present in the $Y$-quadrature statistics as the Hamiltonian breaks the symmetry between $X$ and $Y$; the odd powers of $s$ in the Taylor expansion~\eqref{eq:17e} imply that near $s=0$, ${\theta}_{Y}(s)$ is asymmetric such that, close to the $s=0$ physical dynamics, the sign of $s$ is relevant.  Evaluating the physical $Y$-quadrature activity  we find ${y}_{s=0}=-2/3$, the absolute value of which corresponds to the value of the physical photon emission rate, ${k}_{s'=0}$, as shown in Fig.~\ref{yandk}.
This result is specific to the system parameters here.  While the choice $\kappa=4\Omega$ was identified as special in Ref.~\cite{Garrahan2010} due to a self similarity of quantum-jump statistics, we find this is also the choice of system parameters where the $Y$-quadrature and quantum-jump activities coincide.  That this relation is found for $Y$-quadratures instead of $X$-quadratures is entirely due to the chosen laser polarisation.

\begin{figure}[h]
\centering
\subfloat[]{\label{fig:2lx}\includegraphics[scale=1]{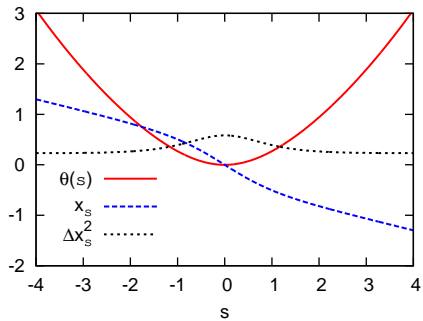}}\\
\subfloat[]{\label{fig:2ly}\includegraphics[scale=1]{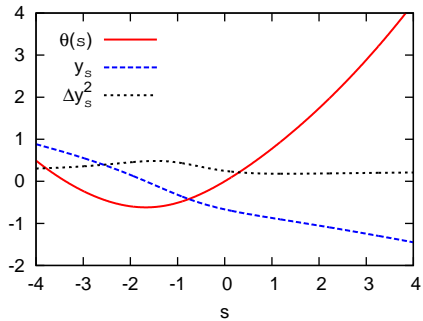}}
\label{fig:4}
\caption{Plots of the LD functions and associated activities and variances for the laser-driven two-level system.  Shown in (a), the LD function for the $X$-quadrature statistics is completely symmetric about $s=0$.  It is also worth noting that as $| s|$ becomes larger, the dynamical variance tends to 1/4, a consequence of the quadratic form of ${\theta}_{X}$ at large $| s|$. In (b) we show that the $Y$-quadrature LD function is now asymmetric close to $s=0$, while at large $|s|$ it adopts the same quadratic form as ${\theta}_{X}(s)$.  Both LD functions exhibit a positive activity for $s<0$ and a negative activity for $s>0$.}
\end{figure}

Switching our focus to the full numerical forms of the LD functions, we show in Figs.~\ref{fig:2lx} and~\ref{fig:2ly} that the minimum of ${\theta}_{Y}$ is shifted away from $s=0$.  However, while $\theta_X(s)$ and $\theta_Y(s)$ are distinct near $s=0$, at very large $|s|$,  ${\theta}_{X}$ and ${\theta}_{Y}$ both tend towards ${s}^{2}/8$.  While the laser breaks the symmetry between $X$- and $Y$-quadratures, the rare trajectories become equivalent again at large $|s|$.  We can understand this as the $s$-field dominating the non-equilibrium dynamics over the self-Hamiltonian of the system.

\begin{figure*}[t]
\includegraphics[scale=1]{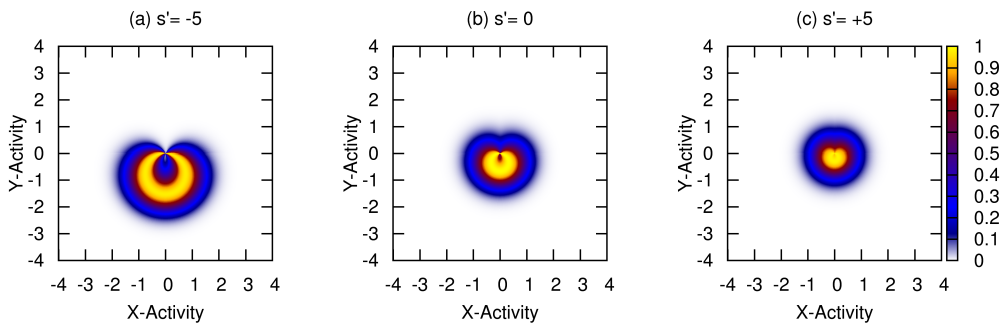}
\caption{Phase-space portraits displaying the marginal probability distributions $e^{-\phi(x^\alpha)}$ for the two-level system at various photon biases $s'$.  We demonstrate that, as we make the system more photon-active, the plot moves away from the origin in the negative $y$-direction.  In (a), ${s}'=-5$ and the center of the plot is (0, -1.534);  in (b), ${s}'=0$ and the plot center is (0, -0.667);  in (c), ${s}'=+5$ and the plot center is at (0, -0.289). Note that although the axes are labelled $x$ and $y$, they are not time extensive.  This is formally equivalent to considering activities when $t=1$.}
\label{fig:5}
\end{figure*}

\begin{figure*}[tb]
\includegraphics[scale=1]{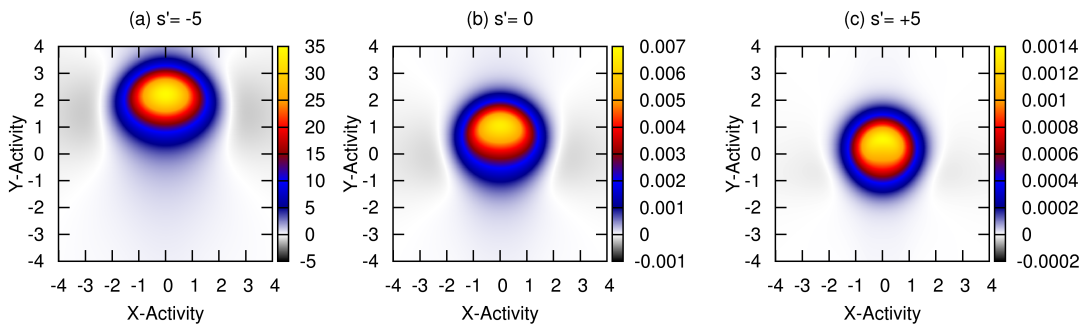}
\caption{Reconstructed Wigner functions for the same system parameters as in Fig.~\ref{fig:5}.  $\alpha$ is incremented by 0.01rad over the range $[0,\pi]$ in the reconstruction.  The values ${s}'= -5$, 0 and $+5$ are plotted in (a), (b) and (c) respectively.  Focus on the central regions where the Wigner function is large: the regions far from the central maxima which take negative values are necessarily sensitive to the choice of $\alpha$-increment used.  We have reproduced the Wigner functions with a broad range of $\alpha$-increments and see no changes to the central regions when choosing increments between 0.01rad and 0.001rad. }
\label{fig:6}
\end{figure*}

We now explore how biasing the photon activities with a counting field ${s}'$ affects the $s=0$ measured quadratures. The dynamics of our doubly biased system are described by the superoperator ${\mathcal{W}}_{s{s}'}$ defined in Eq.~\eqref{eq:16e}.  We diagonalise this superoperator to find the LD function ${\theta}_{K,X^\alpha}(s,{s}')$ and evaluate the quadrature activities by taking the derivative with respect to $s$ at $s = 0$. We look far into both the photon inactive and active regimes discussed in Sec.~\ref{sec:sensemble} by choosing photon counting fields ${s}'=5$ and $-5$ respectively.  To aid visualisation of how the biasing affects the light, we construct a time-independent phase portrait of the light and find the corresponding Wigner distribution, which we define in Appendix~\ref{sec:radon}. To construct our phase portrait we determine the LD function for equally spaced values of $\alpha$ in the range [0,$\pi$] and use the Legendre transform to calculate $e^{-\phi(x^\alpha)}$, from Eq.~\eqref{eq:8e}.  We plot these probability distributions in an activity phase space using coordinates ${x}_{s=0}$ and ${y}_{s=0}$ (hereafter we drop the $s=0$ subscript).  These plots, or portraits, illustrate the set of marginal distributions for all $\alpha$: the marginal probability distributions correspond to cuts through the origin at angle $\alpha$ to the abscissa.  Our results are shown in Fig.~\ref{fig:5}. Using the values $e^{-\phi(x^\alpha)}$ at each ${s}'$, we also construct the Wigner distribution, shown in Fig.~\ref{fig:6}, with a numerical implementation of the inverse Wigner transform.


Figure~\ref{fig:5} shows that the physical dynamics have a ``heart-shaped'' portrait centred on $\left(x,y\right)=\left(0,-2/3\right)$. Although it may appear as though there are bi-modal regions, all $e^{-\phi(x^\alpha)}$ are in fact non-Gaussian unimodal distributions where, as $\alpha$ increases from 0 to $\pi/2$, the mean shifts from 0 to $-2/3$. Interestingly the photon inactive system appears more like a vacuum as it is centred closer to the origin at $(x,y)\simeq(0,-0.289)$. However, although this is a photon-inactive region, there is always a finite probability of quantum jump provided $s$ is finite.
In a complementary fashion the $s'=-5$ portrait ``blows out'' as the centre of each $e^{-\phi(x^\alpha)}$ moves rapidly away from zero, when $\alpha$ increases from $0$ to $\pi/2$. 
  Corresponding features are shown in the Wigner functions plotted in Fig.~\ref{fig:6}.  The Wigner function becomes more concentrated about the origin as we enter the photon-number inactive region, confirming the notion that the output field becomes more vacuum-like as we proceed to more positive ${s}'$.

\section{Driven three-level open systems}
\label{sec:level3}

\begin{figure}[h]
\centering
\includegraphics[height=2cm]{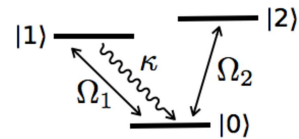}
\caption{A three-level system coupled to a vacuum reservoir and driven by two resonant lasers with Rabi frequencies ${\Omega}_{1}$ and ${\Omega}_{2}$. When ${\Omega}_{1} \gg{\Omega}_{2}$ the photon emission trajectories are intermittent;  the $\ket{1}\rightarrow\ket{0}$ transition is the active or light line transition while $\ket{2}$ is an inactive level.}
\label{fig:3lev}
\end{figure}

We now turn our attention to a more complicated quantum optical system, the three-level system, depicted in Fig.~\ref{fig:3lev}.  We drive this system by two resonant lasers with Rabi frequencies ${\Omega}_{1}$ and ${\Omega}_{2}$ between the $\ket{0}$ level and upper levels $\ket{1}$ and $\ket{2}$ respectively. When ${\Omega}_{1} \gg{\Omega}_{2}$ the photon emission trajectories are intermittent~\cite{Garrahan2010,Plenio1998,barkai2004theory}; the $\ket{1}\rightarrow\ket{0}$ transition is the active or light line transition while the $\ket{2}$ is an inactive level.  Previous studies~\cite{Garrahan2010} found that the ${k}_{s'}$ exhibits a dynamical crossover at $s'=0$. The crossover is from an active phase $s'<0$, where the dynamics are dominated by the $\ket{1}$ and $\ket{0}$ levels, to an inactive phase $s'>0$, where the system dynamics are dominated by long periods where the $\ket{2}$ level is occupied. With this in mind we will examine firstly the quadrature trajectories, then proceed to examine the effects of double biasing and examine the phase portraits of the system.

Considering the three-level system driven by two lasers, the generalized superoperator ${\mathcal{W}}_{s}$ is given in Eq.~\eqref{eq:10e}, still with one set of Lindblad operators $L$ and $L^\dag$.  In this system,  $L=\sqrt{\kappa}\ket{0}\bra{1}$ and there are no other Lindblad operators because of the null decay rate between $\ket{2}$ and $\ket{0}$, as illustrated in Fig.~\ref{fig:3lev}.  However our Hamiltonian for this system is slightly different:
\vspace*{-5mm}
\begin{center}
\begin{equation}
H = \sum_{i=1}^{2}{\Omega}_{i}\left({c}_{i}+{c}_{i}^{\dag}\right),
\end{equation}
\end{center}
where ${c}_{i}\equiv \ket{0}\bra{i}$ and ${c}_{i}^{\dag}\equiv\ket{i}\bra{0}$. When ${\Omega}_{1}\gg{\Omega}_{2}$ typical photon emissions are intermittent displaying both ``bright" and ``dark" periods~\cite{Plenio1998,barkai2004theory}. The LD function $\theta\left(s\right)$ for each $\alpha$ is obtained by direct diagonalization of ${\mathcal{W}}_{s}$ and we examine specifically the case $\kappa=4{\Omega}_{1}$ and ${\Omega}_{2}={{\Omega}_{1}}/{10}$.  

\begin{figure}[h]
\centering
\subfloat[]{\label{fig:3lx}\includegraphics[scale=1]{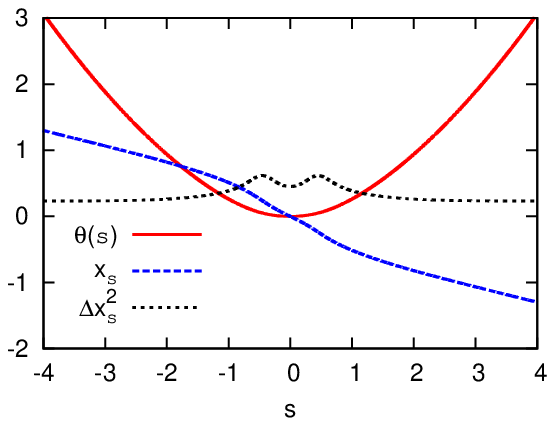}}\\
\subfloat[]{\label{fig:3ly}\includegraphics[scale=1]{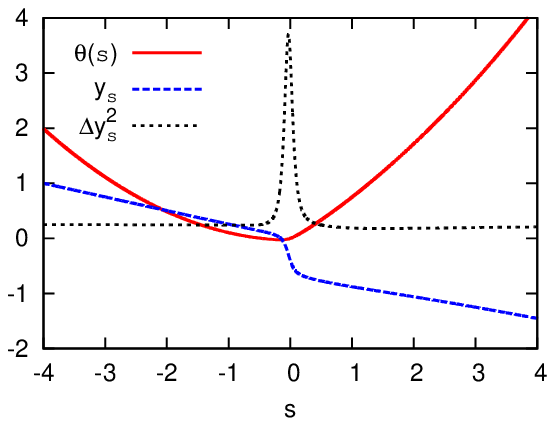}}
\caption{(a) The LD function, activities and dynamical variances for the $X$-quadrature statistics of the three-level system.  This is completely symmetric about $s=0$, but it is not the same as the two-level LD function except in the large $| s|$ limit. This can be seen in the dynamical variance, which tends to 1/4 in this limit, a consequence of the quadratic form of ${\theta}_{X}$ at large $| s|$. (b) The corresponding plots for $Y$-quadratures for the three-level system are shown.  The activity exhibits a marked crossover around $s=0$, which appears as a rounded step.  At the crossover, the corresponding dynamical variance is shown to have a sharp peak.}
\label{fig:8}
\end{figure}

We first study the ensembles of trajectories with $X$-quadrature biases, shown in Fig.~\ref{fig:8}(a).  Here,  ${\theta}_{X}\left(s\right)$ is symmetric in $s$ and exhibits no sharp crossovers.  On the other hand, the ensemble of $Y$-quadratures exhibits a sharp peak in its dynamical variance as we cross $s=0$, as shown in Fig.~\ref{fig:8}(b). This peak corresponds to a crossover between two distinct dynamical phases corresponding to the changeover of the behaviour of $y(s)$ in this region.

In the photon emission study in Ref.~\cite{Garrahan2010}, the crossover in ${k}_{s}'$ was attributed to a change in the effective behaviour where the \emph{active} side where $s'<0$ was argued to be similar to a two-level system.  In the \emph{inactive} phase found for $s'>0$, the $\ket{2}$ level is occupied for long time periods such that few photon emissions occur.  We attribute the crossover observed in $y_s$ to the change in dynamics associated with the crossover in photon emission discussed in Ref.~\cite{Garrahan2010}. To support this claim we examine phase-space portraits of the system and the effects of the $s'$-bias on these portraits, by solving the doubly-biased master equation~\eqref{eq:10e}.

\begin{figure*}[htb]
\includegraphics[scale=1]{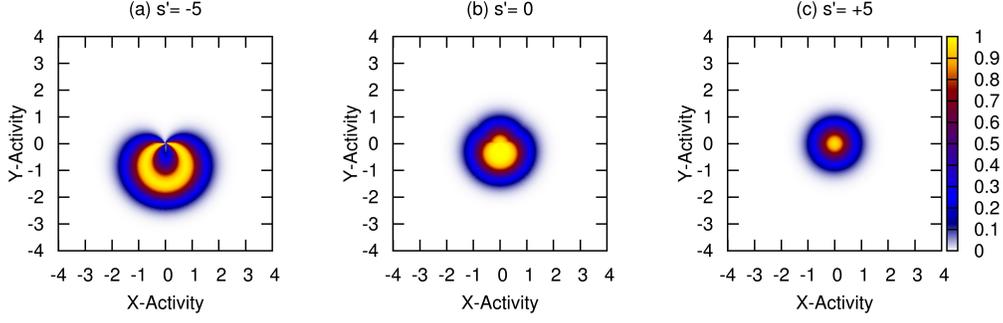}
\caption{Phase-space portraits of the marginal probability distributions $e^{-\phi(x^\alpha)}$ at various photon biases $s'$ for the three-level system.  As we make the system more photon-active the plot moves away from the origin in the negative $y$-direction. In (a), ${s}'=-5$ and the center of the plot is (0, -1.534); in (b), ${s}'=0$ and the plot center is (0, -0.4) and in (c), ${s}'=+5$ and the plot center is at (0,0).}
\label{fig:9}
\end{figure*}

\begin{figure}[h!]
\subfloat[]{\label{c1}\includegraphics[scale= 1]{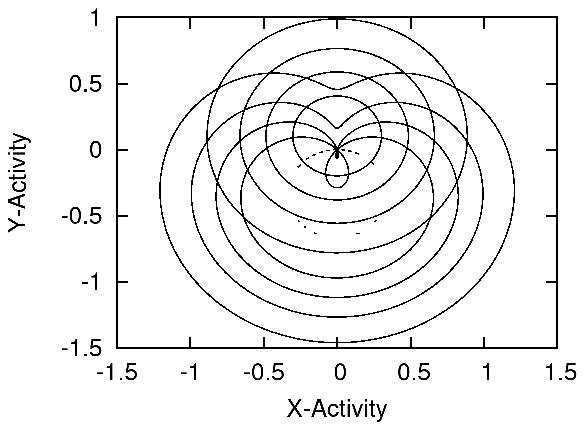}}\\
\subfloat[]{\label{c2}\includegraphics[scale= 1]{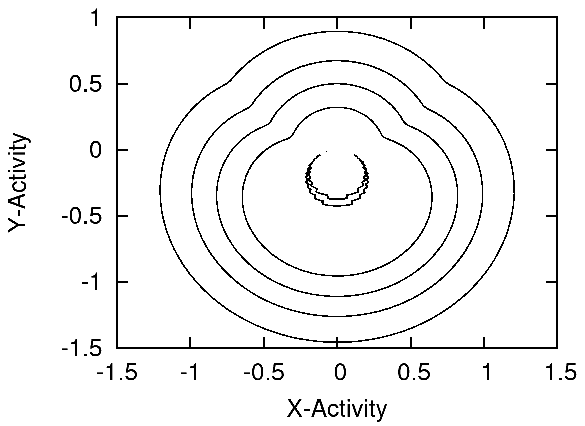}}
\caption{(a) A contour plot of the, ${s}'=0$, two-level system density plot plus the three-level system, ${s}'=+5$, (effectively a one-level system due to inactivity) density plot with its center shifted by 0.1 along the $y$-direction. (b) The contour plot of the three-level ${s}'=0$ marginals.  It is very similar to (a) indicating the three-level system's physical dynamics may be considered as being primarily composed of a two-level system plus an inactive two-level system.  However, the shifts introduced in (a) and slight differences between (a) and (b) are an indication of some interference between these two pictures.}
\label{fig:10}
\end{figure}

First we look at the photon-inactive dynamics with ${s}'=-5$.  The phase portraits and Wigner functions are shown Fig.~\ref{fig:9}(a): these appear identical to the two-level system biased by the same $s'$-field, shown in Fig.~\ref{fig:5}(a). Moving through to ${s}'=+5$, shown in Fig.~\ref{fig:9}(c), the phase portrait appears almost like a vacuum.  This is not a true vacuum; both photon activity and $y_s$ for this portrait are small but non-zero. 
However the forms of the portrait and Wigner function do suggest that for large $s'$, the $\ket{2}$ state is occupied for a large fraction of the time.  We illustrate our finding more clearly in Figs~\ref{c1} and~\ref{c2}, where we show how a contour plot of the $s=0$ two-level portrait with three-level shifted $s=+5$ contours closely resembles the three-level system $s=0$ plot.  This demonstrates that the physical dynamics of the three-level system is effectively made up of an active two-level system plus an inactive two-level system (this is where the dynamics are dominated by occupations of the $\ket{2}$ state)  with some ``mixing'' of the two. 

\begin{figure}[h!]
\centering
\includegraphics[scale=1]{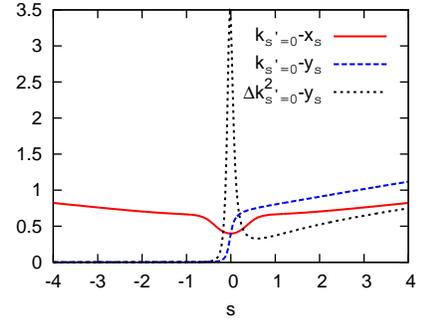}
\caption{Plots of the quantum-jump activity $k_{s'}$, for different quadrature biases $s$ in the three-level system.  The jump activity grows as we bias the $X$-quadrature regardless of sign of $s$.  However the photon count exhibits a dynamical crossover in the rare $Y$-trajectories about $s=0$, and shows that the sign of $s$ determines the behaviour of ${k}_{{s}'}$.  Furthermore the photon variance in these quadratures also indicates the inactive/active phase crossover, with a peak at $s=0$ correlating with the dynamical phase transition at this point}.
\label{fig:11}
\end{figure}

To determine whether or not it is this ${k}_{s'}$ crossover which is responsible for the crossover in $y_s$ we examine the statistics of typical jump trajectories for rare quadrature trajectories. We repeat the procedure outlined in Sec.~\ref{sec:s2} but now we first bias quadratures and then count quantum jumps by introducing the doubly-biased density matrix
\vspace*{-5mm}
\begin{center}
\begin{equation}
{\rho}^{{s}'s}(t)={\text{Tr}}_{\text{res}}\left({{V}^{s}}^{\frac{1}{2}}{V}^{{s}'}{{V}^{s}}^{\frac{1}{2}}\rho \right)\,.
\end{equation}
\end{center}
Then, we generate a new master equation
\vspace*{-5mm}
\begin{center}
\begin{eqnarray}
&&\dot{{\rho}}^{{s}'s}(t)=\mathcal{L}\left({\rho}^{{s}'s}\right) +\left({e}^{-s'}-1\right)\kappa {c}_{1}{\rho}^{{s}'s}{c}^{\dag}_{1}\\ 
&&+\frac{{s}^{2}}{8}{\rho}^{{s}'s}-\frac{s\sqrt{\kappa}\left({e}^{-{s}'}+1\right)}{4}\left({e}^{-i\alpha}{c}_{1}{\rho}^{{s}'s}+{e}^{i\alpha}{\rho}^{{s}'s}{c}^{\dag}_{1}\right). \nonumber
\end{eqnarray}
\end{center}
Solving for the LD function, ${\theta}_{{X}^{\alpha},K}(s',s)$ for both the $X$- and $Y$- quadratures we find $k_{s'=0}$ for various quadrature biases $s$. The typical activity $k$ shows no sharp features as we bias the $X$-quadrature, irrespective of the sign of $s$;.  However, if we bias the $Y$-quadrature, the activity $k$ and variance $\Delta k^2$ show photon-number inactivity when $s<0$.  That is to say, more positive $y_s$ leads to smaller ${k}$.  This is demonstrated in Fig.~\ref{fig:11}. This supports our previous assertion that the activities ${y}_{s}$ and ${k}_{{s}'}$ may be used as equivalent order parameters in this system.
\begin{figure}[h!]
\centering
\includegraphics[scale=1]{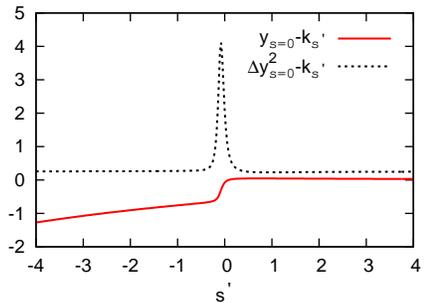}
\caption{Plots of the activity $y_{s=0}$ as a function of photon bias $s'$ in the three-level system.  $y_{s=0}$ exhibits a crossover at $s'=0$, confirming that while ${k}_{{s}'}$ varies with $s$ in a highly correlated way,  ${y}_{s}$ is similarly correlated with ${s}'$.}
\label{fig:12}
\end{figure}
For completeness, we show in Fig.~\ref{fig:12} how $y_{s=0}$ varies as a function of $s'$.  Surprisingly, we find the same correlation at $s=0$.  This result is surprising because, although no inactive phase exists for ${y}_{s}$, it is still a valid dynamical order parameter due to its connection to the quantum-jump activity: for ${y}_{s=0}$, the region where ${s}'>0$ corresponds to the active quantum-jump phase and ${s}'<0$ corresponds to the inactive phase.  This connection between $y_{s=0}$ and ${k}_{{s}'}$ also exists in the two-level system studied in Sec.~\ref{sec:level2}, and may be a trait present in many systems.

To conclude, the finding that quadrature activities may encode the same dynamical statistics as the photon activity is quite surprising,  given the non-commuting nature of the operators.  We next ask whether there are systems where a dynamical crossover is only visible through the quadrature activity.  We will discuss this question in Secs.~\ref{sec:2L2C} and~\ref{sec:MICRO}.

\section{Two Coupled Two-Level Systems}
\label{sec:2L2C}

\begin{figure}[h!]
\centering
\includegraphics[height=3cm]{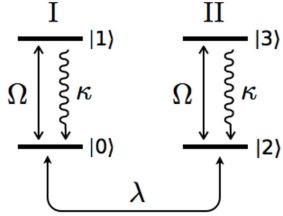}
\caption{A schematic diagram of two weakly-coupled two-level systems driven by lasers of opposite circular polarisation.  The coupling between the ground states is neglected in our case.}
\label{fig:13}
\end{figure}

We extend our study to a pair of coupled two-level systems driven by lasers with different polarisations, depicted in Fig.~\ref{fig:13}.  In the previous two sections, we have identified that crossovers in quadrature activity mirror crossovers in the number of emitted photons, showing an equivalence between quadrature activity and photon activity as dynamical order parameters.  Here we show that crossovers can occur in the quadrature activity and not in the photon activity, emphasising the use of quadrature activity as a dynamical order parameter in its own right.  

We consider two weakly coupled two-level systems as shown in Fig.~\ref{fig:13}. There are now two Lindblad operators associated with each two-level system: ${L}_{I}=\sqrt{\kappa}{c}_{I}$ and ${L}_{II}=\sqrt{\kappa}{c}_{II}$, where these operators are the ladder operators of each two-level system, I and II respectively. These two-level systems have identical Rabi frequencies and decay rates.  As previously, we choose $\kappa =4\Omega$, but distinguish the two subsystems by the polarisation of the driving laser. Subsystem I is driven by a ${\sigma}_{x}-{\sigma}_{y}$ polarized laser, while subsystem II is driven by a ${\sigma}_{x}+{\sigma}_{y}$ polarized laser.  The Hamiltonian is
\vspace*{-5mm}
\begin{center}
\begin{eqnarray}\label{eq:4lH}
H &=& \Omega\left({c}_{I}+{c}^{\dag}_{I}-i{c}_{I}+i{c}^{\dag}_{I}+{c}^{\dag}_{II}+{c}_{II}+i{c}_{II}-i{c}^{\dag}_{II}\right) \nonumber \\
&+&\lambda\left(\ket{0}\bra{2}+\ket{2}\bra{0}\right) .
\end{eqnarray}
\end{center}
This model allows quantum coherence between the two-level systems to be preserved.  We choose the coupling $\lambda$ $\ll \Omega$, with $\lambda = \Omega/10$ in the presented results.  The weak coupling along with the similar physical properties of two subsystems is such that dynamical crossovers in quantum jump trajectories of this system can occur.  However the different polarizations of the lasers driving each subsystem affords the possibility of a transition in quadrature trajectories. 

\begin{figure}[h!]
\centering
\includegraphics[scale=1]{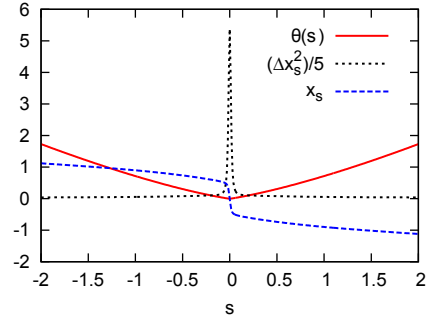}
\caption{Plots of the $X$-quadrature LD function as well as the activity and dynamical variance, for the four-level system depicted in Fig.~\ref{fig:13}.  Note that the dynamical variance has been rescaled to be a factor of 5 smaller.}
\label{fig:14}
\end{figure}

While the statistics of the $Y$-quadrature is featureless, Fig.~\ref{fig:14} shows that the $X$-quadrature displays a crossover around $s=0$. This crossover is smeared out if we increase $\lambda$ and becomes sharper if we decrease $\lambda$. The $s<0$ region has a positive $X$-activity; this is due to light emitted from subsystem I.  When $s>0$, the light is emitted from subsystem II and the $X$-activity is negative. We expect that the $s=0$ behaviour is a combination of these two distinct dynamical phases and, to demonstrate this is the case, we construct time-independent probability distributions, or phase portraits, for each dynamical regime.

\begin{figure*}[tb]
\includegraphics[scale=1]{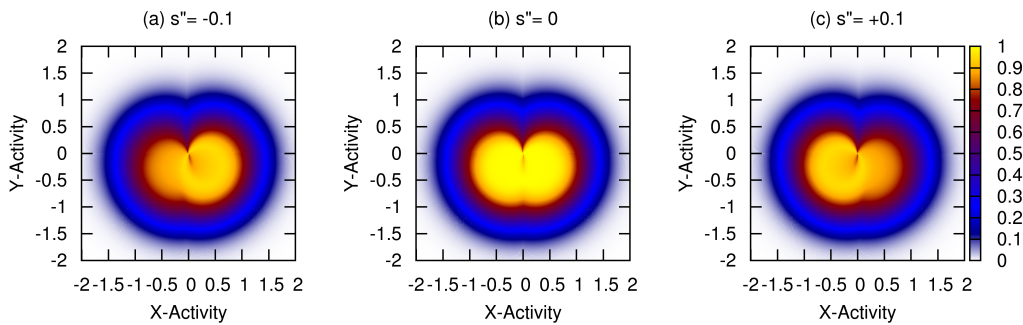}
\caption{Phase-space portraits for the two coupled two-level systems.  The $X$-quadrature bias in (a), (b) and (c) is $s''=-0.1$, 0 and $+0.1$  respectively. In (a) and (c) the probability distributions are more concentrated about $x>0$ and $x<0$ respectively, whereas at $s''=0$ they are even functions of $x_s$. These plots are indicative of the crossover in the $X$-activity at $s=0$ and highlight the less dramatic phase changes we may see using this order parameter as opposed to the jump activity.}
\label{fig:15}
\end{figure*}
We wish to construct the marginal distributions $e^{-\phi(x^\alpha)}$ for rare $X$-quadrature trajectories. We will bias the $X$-quadrature statistics using a field denoted $s''$ and measure the typical $X^\alpha$ for all $\alpha$ using the conjugate field $s$. To this end we apply the double biasing scheme as in Eq.~\eqref{eq:14e} where instead of biasing with the jump operator and measuring ${X}^{\alpha}$, or vice-versa, we now bias the $X$-quadrature.  This leads to an equivalent doubly-biased master equation
\vspace*{-5mm}
\begin{center}
\begin{align}
&\dot{{\rho}}^{s{s}''}(t)=\mathcal{L}\left({\rho}^{s{s}''}\right)+{\rho}^{s{s}''}\frac{s{s}''}{4}\cos\left(\alpha\right)\nonumber\\ &-\frac{s\sqrt{\kappa}}{2}\sum_{i}{\left({e}^{-i\alpha}{c}_{i}{\rho}^{s{s}''}+{e}^{i\alpha}{\rho}^{s{s}''}{c}^{\dag}_{i}\right)} \nonumber\\
&-\frac{{s}''\sqrt{\kappa}}{2}\sum_{i}{\left({c}_{i}{\rho}^{s{s}''}+{\rho}^{s{s}''}{c}^{\dag}_{i}\right)} \nonumber\\
&+\frac{{s''}^{2}}{8} {\rho}^{s{s}''} + \frac{{s}^{2}}{8}{\rho}^{s{s}''},
\end{align}
\end{center}
where the sum is over the two subsystems. Using ${s}''$ we bias the system to rare $X$-quadrature trajectories and we look at derivatives of the LD function with respect to $s$ in the limit $s\rightarrow 0$ to measure the quadratures of these biased trajectories.  Solving for the probability distributions, for ${s}''=-0.1$, 0 and $+0.1$, we find that the behaviour of typical ($s=0$) is made of two distinct dynamical phases, as is clear in Fig.~\ref{fig:15}.

This crossover of emission from one subsystem to another is not observable in the jump trajectories and highlights that quadrature activities are not just equivalent to the jump activity as a dynamical order parameter. These quadrature activities may reveal extra dynamical phases which are not visible simply by counting photons.  

\section{Micromaser}
\label{sec:MICRO}
The final example we present is a many-body problem consisting of a set of two-level atoms interacting with a single cavity mode, called a micromaser~\cite{Scully1997,Haroche2006,Walther2006,Filipowicz1,Garrahan2011}. We begin with a brief description of this problem before providing a mean-field treatment to determine the LD function. We then proceed to determine the full LD function with exact numerical diagonalisation and produce quadrature activity phase diagrams for the system.  Finally, we examine the doubly-biased trajectory spectrum of the system.  We then summarise our findings, before moving on to present our conclusions in the final section of this work, Sec.~\ref{sec:Conc}.

The micromaser is a single-mode resonant cavity coupled to a thermal bath and pumped by excited two-level atoms which pass through the cavity.  We denote the total number of atoms which pass through the cavity divided by the cavity lifetime by ${N}_{ex}$. The steady-state cavity occupation distribution can change from uni-modal to bi-modal, depending on ${N}_{ex}$ and the atom-cavity coupling.   Here, we fix $N_{ex}=100$.   As we increase the coupling between the cavity mode and the atoms, we reach many points where the steady state cavity occupation number exhibits a bistability:  at these points, a small increase in the coupling leads to a crossover in occupation number where the occupation number changes dramatically.  Previously this bistability in the state of the cavity was shown to have an equivalent dynamical bistability in quantum-jump trajectories in Ref.~\cite{Garrahan2011}.  A rich dynamical phase structure was found, where the number of atoms which have changed state while traversing the cavity was used to quantify the dynamical activity.

We now turn to the model for the micromaser.  Tracing out the atom and bath degrees of freedom, one obtains the superoperator, $\mathcal{W}$, which contains no coherent evolution term and is of Lindblad form that is of the form in Eq.~\eqref{eq:3e} with $H=0$ and four pairs of Lindblad operators. There are two associated with the atom-cavity interaction
\begin{eqnarray}
 {L}_{1}&=& \sqrt{{N}_{ex}} {a}^{\dag}  \frac{\sin\left (\phi \sqrt{a{a}^{\dag}} \right ) }{\sqrt{a{a}^{\dag}} } \\
{L}_{2}&=&\sqrt{{N}_{ex}}\cos\left(\phi\sqrt{a{a}^{\dag}}\right)
\end{eqnarray}
and two result from the cavity-bath interaction, 
\begin{eqnarray}
{L}_{3}&=&\sqrt{\nu+1}\,a \\
{L}_{4}&=&\sqrt{\nu}{a}^{\dag}\,.
\end{eqnarray} 
Here $\phi$ is the accumulated Rabi frequency which encodes the atom-cavity interaction, $a^\dag$ and ${a}$ are the cavity raising and lowering operators respectively.   Choosing a zero-temperature bath, we will set the thermal occupation number of the bath $\nu=0$.

We study the quadratures of the light emitted into to the bath and therefore deform the superoperator by the non-equilibrium quadrature field to
\vspace*{-5mm}
\begin{center}
\begin{equation}\label{eq:19e}
{\mathcal{W}}_{s}\left({\rho}^{s}\right) = \mathcal{W}\left({\rho}^{s}\right) -\frac{s}{2}\left({e}^{-i\alpha}{L}_{3}{\rho}^{s} +{e}^{i\alpha}{L}_{3}^{\dag}{\rho}^{s}\right) +\frac{{s}^2}{8}{\rho}^{s} . 
\end{equation}
\end{center}
For this system, as the normal dynamics of the system are purely Lindbladian in nature, the LD function for the quadratures is identical for all $\alpha$.  Therefore in the following discussion we will focus solely on the $X$-quadrature (where $\alpha = 0$) from here on. 

\subsection{Mean-Field Approximation}
\begin{figure}[h!]
\centering
\includegraphics[scale=1.25]{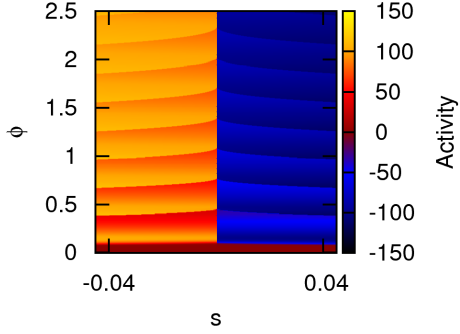}
\caption{Quadrature activity in the micromaser from mean-field theory.  We find multiple first order transition lines in the activity either side of $s=0$.  Also we note some bending of these transition lines occurs as they approach $s=0$. The form of the mean-field theory is very similar to one obtained for the ``atom" counting case of Ref.~\cite{Garrahan2011}, highlighting the connection between transitions in quadrature and jump-activities.}
\label{fig:16}
\end{figure}

Although a full analytic solution for the $s=0$ steady state density operator is well known~\cite{Filipowicz1,Garrahan2011}, away from $s=0$ it is not easily generalised. In a study of quantum-jump trajectories of this system~\cite{Garrahan2011} progress was made by assuming the eigenmatrix of an $s'$-dependent superoperator was diagonal in ${a}^{\dag}a$.  This approach demonstrated that the cavity pump rate controlled the properties of the coexistence line at $s'=0$ and the findings agreed well with exact diagonalisation results.  Mulitple first order transition lines were observed in the photon active region where $s'<0$, and a single transition was observed in the photon inactive regime where $s'>0$, along with a critical point at $s'\approx0$ and $\phi\approx0.1$.  In this work, the effects of quadrature biasing with $s$ using ${\mathcal{W}}_{s}$ in Eq.~\ref{eq:19e} leads to eigenmatrices which  are not diagonal in ${a}^{\dag}a$ for general $s$. However close to $s=0$ we approximate the non-diagonal term, $-\frac{s}{2}\left({L}_{3}{\rho}^{s} +{L}_{3}^{\dag}{\rho}^{s}\right)$, with a diagonal one given by

\vspace*{-5mm}
\begin{center}
\begin{align}
&-\frac{s}{2}\left(a{\rho}^{s}+{\rho}^{s}{a}^{\dag}\right)  \nonumber \\ 
&=-a{\rho}^{s}{a}^{\dag}+\left(a-\frac{s}{2}\right){\rho}^{s}\left({a}^{\dag}-\frac{s}{2}\right) -\frac{s^2}{4} \rho^s\nonumber\\ 
&\approx -a{\rho}^{s}{a}^{\dag}+{e}^{\left |s\right |}a{\rho}^{s}{a}^{\dag}.
\end{align}
\end{center}
Here we have assumed that $a$ and ${a}^{\dag}$ are approximately $\sqrt{n}$, where $n$ is the cavity occupation number.  Thus for $s\ll 1$ we may approximate the non-diagonal piece by an exponential and the appearance of $|s|$ reflects the fact that $\theta_X(s)$ must be an even function of $s$. 

Through this crude approximation we have restricted our analysis to density operators diagonal in the number basis, and so the generalized quantum master equation reduces to an operator
\vspace*{-5mm}
\begin{center}
\begin{align}\label{eq:20e}
{\mathcal{W}}_{s}\rightarrow {W}_{s} &={N}_{ex}{a}^{\dag}\frac{\sin^2\left(\phi\sqrt{{a}^{\dag}a+1}\right)}{\sqrt{{a}^{\dag}a+1}} \nonumber\\ &-{N}_{ex}\sin^2\left(\phi\sqrt{{a}^{\dag}a+1}\right) -{a}^{\dag}a\nonumber\\ &+{e}^{\left |s\right |}\sqrt{{a}^{\dag}a+1}\,\,a - \frac{{s}^{2}}{8}.
\end{align}
\end{center}
A variational approach to calculating the largest eigenvalue of ${W}_{s}$ in Eq.~\eqref{eq:20e} can be constructed using a coherent state ansatz.  This amounts to setting $a={e}^{i\gamma}\sqrt{n}$ and ${a}^{\dag}={e}^{-i\gamma}\sqrt{n}$ and solving the corresponding Euler-Lagrange equations, $\partial{W}_{s}/\partial\gamma = 0$ and $\partial{W}_{s}/\partial n = 0$. Solving the first equation one obtains,

\vspace*{-5mm}
\begin{center}
\begin{align}
&a =\sqrt{n}{\left({{e}^{-\left|s\right |}{N}_{ex}}\frac{\sin^2\left(\phi\sqrt{n+1}\right)}{{n+1}}\right)}^{1/2}\nonumber \\
&{a}^{\dag}=\sqrt{n}{\left({{e}^{-\left|s\right|}{N}_{ex}}\frac{\sin^2\left(\phi\sqrt{n+1}\right)}{{n+1}}\right)}^{-1/2}.
\end{align}
\end{center}
Substituting these into ${W}_{s}$~\eqref{eq:20e} one obtains a variational ``free energy", ${\mathcal{F}}_{s}\left(n\right)$, whose minimum with respect to $n$ yields an estimate for the LD function,
\vspace*{-5mm}
\begin{center}
\begin{equation}
\theta_X(s)\approx -\text{min}_{n}{\mathcal{F}}_{s}\left(n\right).
\end{equation}
\end{center}

This minimization process was performed numerically and the result is displayed in Fig.~\ref{fig:16}. Minimization reveals that multiple first-order transitions occur in both the cavity occupation number and quadrature activity.  These transition lines begin to bend as they approach $s=0$. The first transition line ends at $s\approx 0$, $\phi \approx 0.1$. This point is the critical point discovered in Ref.~\cite{Garrahan2011}, which controls the photon number dynamics. Here it is much more masked, even within this crude diagonal approximation, than in the study of jump-trajectories of the atoms presented in Ref.~\cite{Garrahan2011}. Within this diagonal approximation, Eq.~\eqref{eq:20e}, there are limitations and further implicit assumptions: the approximation becomes less accurate at larger $\phi$ where non-linearities in the operator Eq.~\eqref{eq:20e} are more prominent. There is also an implicit assumption of normal ordering and that the averages of products of the raising and lowering operators may be replaced by products of their averages. Despite ignoring the off-diagonal behaviour ${\mathcal{W}}_{s}$, this free energy predicts multiple first order transitions in the quadrature activity and, moreover, that these transitions are due to transitions in the occupation number, $n$, of the cavity.

\subsection{Full numerical diagonalization}
\label{sec:FND}
\begin{figure*}[htb]
\includegraphics[scale=1.25]{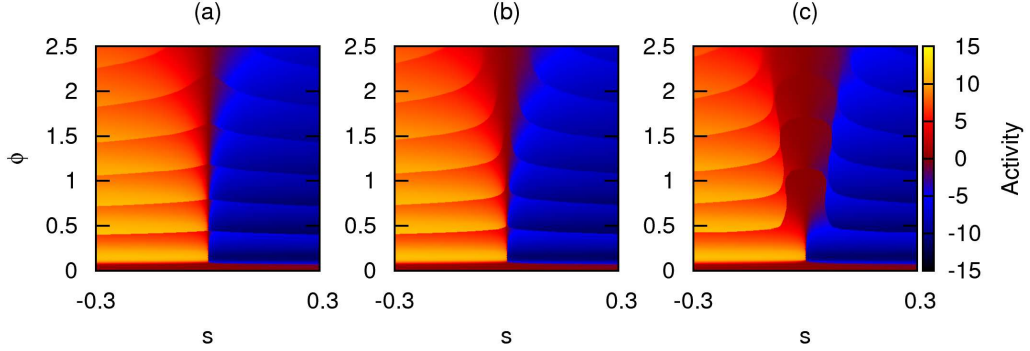}
\caption{Quadrature activity phase diagrams in quantum-jump biased systems with $s'= -0.005$, 0 and $+0.005$ in plots (a), (b) and (c) respectively. In all cases the activity shows first order transition lines as we change $\phi$.  Comparing (b) with the mean field theory, agreement exists up to $\phi\approx 0.7$,  however there is significant bending of these lines as we approach the normal system dynamics. The degree of this bending is linked with the jump activity: in the active phase (a), the transition lines bend less approaching $s=0$, while in the inactive jump phase, (c), the bending is more prominent.}
\label{fig:17}
\end{figure*}

\begin{figure*}[htb]
\includegraphics[scale=1.25]{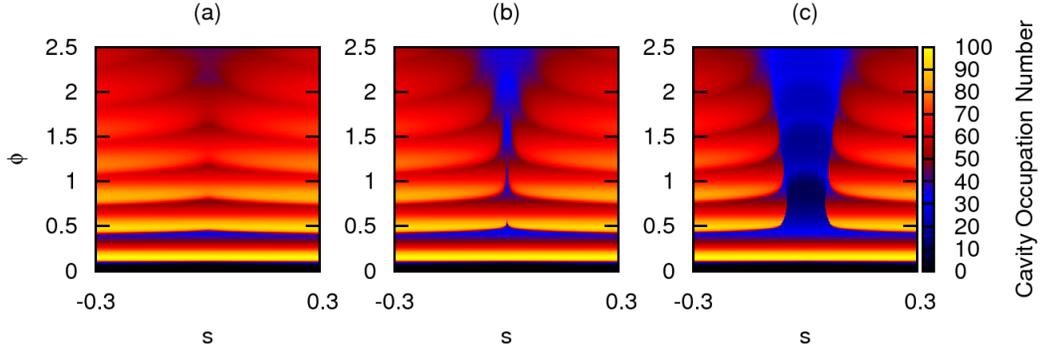}
\caption{Cavity occupation number for double counting cases with jump bias $s'=-0.005$, 0 and $+0.005$ in plots (a), (b) and (c) respectively. In the jump-active case, (a), the reduced bending of the transition lines compared with (b) correlates with an increased cavity occupation which remains larger than the typical value away from $s=0$. Conversely the increased bending in (c) compared with in (b) is associated with larger regions away from $s=0$ attaining typical ($s=0$) cavity occupation numbers rapidly as we increase $\phi$.}
\label{fig:18}
\end{figure*}

We now examine the generalized master equation in order to determine the exact form of the LD function. The generalized master equation may be expressed in terms of a matrix and diagonalized numerically.  The method we employ is similar to that in Ref.~\cite{Genway1}, and outlined in Appendix~\ref{sec:DIAG}.  The full phase diagram is shown in Figure~\ref{fig:17}(b), where we see that the critical point previously discussed manifests itself as the only point where the first order transition lines accumulate at $s=0$. Beyond $\phi\approx0.1$ the transition lines begin to bend in such a manner as not to accumulate at the $s=0$ line.  It is clear however that the mean-field theory in Fig.~\ref{fig:16} predicts values of $\phi$ where the transitions occur up to $\phi\approx 0.7$.  These transitions correlate with the cavity-occupation plot shown in Fig.~\ref{fig:18}(b), demonstrating that these first order transitions in quadrature activity are also associated with the bistability of the cavity photon number~\cite{Garrahan2011}.

We now examine the statistics of double biasing, using the field $s'$ to place the system in higher or lower photon number state and then examining the typical quadrature realisations.  Notice that the transition lines shift such that they either accumulate more at $s=0$ or bend away further before reaching $s=0$ when placing the system in a high ($s'<0$) or low ($s'>0$) cavity number respectively.  Placing the system in a $k$-active phase removes lower-cavity-number regions near $s=0$, visible in Fig.~\ref{fig:18}(b), causing the quadrature activity lines to meet at $s=0$.  Conversely, biasing the dynamics towards $k$-inactive trajectories ensures that the cavity occupation is small unless a large $s$-bias is applied.  This can be seen in the formation of a band around $s=0$ where both the quadrature activity (Fig.~\ref{fig:17}) and cavity occupation (Fig.~\ref{fig:18}) are small. 

Although sharp crossovers occur in the quadrature activity, corresponding to crossovers in the cavity occupation, it is not clear that the $s=0$ dynamics are made up of two distinct phases. These phases are only apparent by examining $k$.  Furthermore, even though the critical point exists in both the $k$ and $x$ phase diagrams, it is much easier to discern in the former. To conclude, we find that quadrature measurements in micromaser can be used to identify dynamical phases, but in sharp contrast to the four-level system in~\ref{sec:2L2C}, the study of quadrature trajectories does not give significant insights into the system dynamics beyond those available by counting photons.

\section{Conclusions}
\label{sec:Conc}
In this paper we developed an $s$-ensemble approach to gain insight into the behaviour of light emitted from several quantum-optical systems.  Previous studies used the $s$-ensemble to describe the dynamics of quantum-jump trajectories and understand the statistics of photon emissions in open quantum systems.  We reformulated the approach in terms of generating functions and characteristic operators, treating the reservoir stochastically. Using this approach we examined the statistics of the quadratures of light emitted from various systems and re-constructed appropriate phase space portraits and Wigner functions which describe the state of the emitted light.  We extended this scheme to examine the light not only emitted during rare quantum-jump trajectories where, for example, the number of emitted photons is larger or smaller than for average dynamics.  We also examined the quantum-jump statistics of the quadrature-biased trajectories, ascertaining the observable photon count when the quadratures of emitted light depart significantly from their dynamical averages.  We used these results to understand whether different dynamical order parameters are correlated, and to select the most appropriate $s$-field and order parameter with which to characterise dynamical phases.

Our results show that even simple quantum-optical models such as two- and three-level systems show interesting features in the space of quadrature trajectories.  For example, for the case of the blinking three-level system, quadrature phase-space portraits indicate that emitted light may be identified with a mixture of the light from two simple active and inactive subsystems, which is consistent with the dynamical coexistence description in terms of number of emitted photons described in Ref.~\cite{Garrahan2010}. For the case of the two-level system, our analysis here shows that the `special point' in parameter space found in Ref.~\cite{Garrahan2010} is also special for another reason: it is the point where one may measure the $Y$-quadrature activity $y_s$ in place of the quantum-jump activity $k_{s'}$ of the physical system, as they are identical in magnitude: for this system ${y}_{s}$ represents an alternative dynamical order parameter, as biasing trajectories towards larger ${y}_{s}$ leads to smaller ${k}_{{s}'=0}$, and vice versa.

Examination of a pair of coupled two-level systems showed that, in some circumstances, the quadrature activity reveals a phase structure which is completely unobservable through photon count statistics.  This highlights the usefulness of studying quadrature trajectories, and illustrates that greater inference about an open system can be obtained by such measurements.  However, it also highlights that the choice of dynamical order parameter is highly system dependent.  We further illustrated the difficulty in the choice of order parameter by studying quadratures in the micromaser.  The phase diagram for the quadrature dynamics is rich and displays many features of the quantum-jump activity phase diagram studied in Ref.~\cite{Garrahan2011}, such as numerous first order transition lines.  However, the lack of direct control over the cavity-state bistability means one cannot simply describe the $s=0$ dynamics in terms of two distinct quadrature phases.

Finally, we emphasize the significance of our results in light of homodyne detection schemes, which allow $X$-quadrature trajectories to be measured.  Furthermore, we note that the $Y$-quadrature statistics discussed are accessible to experimental measurement with a suitable modification to the driving laser polarisation.  We also add that quantum tomographic techniques allow reconstruction of the Wigner functions of such systems.  We further note that the methodology presented in this work can be extended to other more complex measurement schemes with other unravellings of the quantum dynamics for different choices of bath measurement operators with both discrete and continuous spectra.

\acknowledgements

We are grateful to Madalin Guta for important discussions.  This work was supported by 
The Leverhulme Trust Grant no.\ F/00114/B6 and EPSRC Grant no.\  EP/I017828/1.


\appendix
\section{Derivation of Quadrature s-Ensemble Master Equation}
\label{sec:Der1}

Using a weakly coupled Markovian reservoir one may consider that the ladder lowering operator of the reservoir acts as a driving field for the time-evolution operator of the total system, and so the effects of the reservoir ladder operator fields permit a description in terms of an input-output formalism~\cite{gardiner2004}. In general, such a formalism works on the foundation that outside the range of interaction, the field which interacts with the system is the sum of the input field, that is the field just prior to interacting with the system, and the output field, the state of the field just after interaction with the system. The reservoir ladder operators, $b(t )$ and ${b}^{\dag} (t' )$, are the fields interacting with the system, the time $t$ is the initial time when the fields interact with the system and, due to our choice of a reservoir which admits a Fock space representation, one may use a stochastic description of the system-plus-bath dynamics. Although the input-output formalism is intuitive, especially if one is working with quantum optics where the input/output fields are manifested as light beams, the stochastic description~\cite{gardiner2004} of the system allows for a concise mathematical representation where the effects of the reservoir on the system are reduced to an It\={o} increments which obey a quantum It\={o} calculus.

Before we proceed with the derivations of the generalized master equation for biased quadrature ensembles, it is necessary to outline the calculus of the reservoir increments, as well as the quantum-jump $dK$ process. Working with an unsqueezed vacuum, the only non-zero combination of $dB(t)$ and $d{B}^{\dag}(t)$ is:
\vspace*{-5mm}
\begin{center}
\begin{equation}\label{eq:4e}
dB(t){dB}^{\dag}\left(t \right)= dt.
\end{equation}
\end{center}
With this It\={o} table we can deduce the appropriate table for the jump trajectory process $dK$,
\vspace*{-5mm}
\begin{center}
\begin{align}\label{eq:13e}
dK(t)dK(t)&=dK(t)\nonumber \\
dB(t)dK(t)&=dB(t)\nonumber\\
dK(t)d{B}^{\dag}(t)&=d{B}^{\dag}(t) \, .
\end{align}
\end{center}
Now, using the definition of ${\rho}^{s}$~\eqref{eq:rhoQ}, we examine the increment of the $s$-biased density matrix
\vspace*{-5mm}
\begin{center}
\begin{equation}\label{eq:I}
d[{\rho}^{s}] = {\text{Tr}}_{\text{Res}}\left(d[{V}_{Q}^{s}\left(t\right)]\rho + {V}_{Q}^{s}\left(t\right)d[\rho] + d[{V}_{Q}^{s}\left(t\right)]d[\rho]\right)\,.
\end{equation}
\end{center}
The first two terms appear in standard calculus while the final term appears due to the stochastic nature of these increments. We have already defined $d\rho$ in Eq.~\eqref{eq:2e}, and so we need to examine the increment of the characteristic operator.  Expanding ${V}_{X^\alpha}^{s}\left(t\right)$, defined in ~\eqref{eq:14e}, and using the calculus set out in Eq.~\eqref{eq:4e} one finds
\vspace*{-5mm}
\begin{center}
\begin{equation}
d[{V}_{X^\alpha}^{s}\left(t\right)] = {V}_{X^\alpha}^{s}\left(t\right)\left(\frac{{s}^{2}}{8}dt -\frac{s}{2}d{X}^{\alpha}\right).
\end{equation}
\end{center}
We may now evaluate the expansion on the right-hand side of~\eqref{eq:I}, term by term, to obtain
\vspace*{-5mm}
\begin{center}
\begin{align}
&{\text{Tr}}_{\text{Res}}\left(d[{V}_{X^\alpha}^{s}\left(t\right)]\rho\right)= \frac{s^{2}}{8}{\rho}^{s}dt, \nonumber \\
&{\text{Tr}}_{\text{Res}}\left({V}_{X^\alpha}^{s}\left(t\right)d[\rho]\right)= \mathcal{L}\left({\rho}^{s}\right)dt,\nonumber \\
&{\text{Tr}}_{\text{Res}}\left(d[{V}_{X^\alpha}^{s}\left(t\right)]d[\rho]\right)=-\sum_{i}\frac{s}{2}\left({e}^{-i\alpha}{L}_{i}{\rho}^{s}+{e}^{i\alpha}{\rho}^{s}{L}_{i}^{\dag}\right)dt, 
\end{align}
\end{center}
where the ${L}_{i}$ are the Lindblad operators of the system coupled to the environment. We can now see the increment equation~\eqref{eq:I} leads to Eq.~\eqref{eq:10e}. 

This procedure may be extended straightforwardly to the doubly-biased cases.  In the case of the biasing photon trajectories and examining the typical quadrature statistics (Eq.~\eqref{eq:15e}) one introduces the jump-trajectory characteristic operator ${V}^{{s}'}_{K}\left(t\right)$ as in Eq.~\eqref{eq:4e} and defines its increment
\vspace*{-5mm}
\begin{center}
\begin{equation}
d[{V}^{{s}'}_{K}\left(t\right)] = {V}^{{s}'}_{K}\left(t\right)\left({e}^{-{s}'}-1\right)dK.
\end{equation}
\end{center}
Now we proceed as before using the It\={o} calculus set out in Eqns.~\eqref{eq:4e} and~\eqref{eq:13e} to arrive at the result Eq.~\eqref{eq:15e}.
Finally we note that, although introduced for a specific set-up, this formalism is not restricted to pure states and allows a freedom of choice for the reservoir provided we stay within the Markovian approximation.  For example, it may be readily generalized to a squeezed thermal reservoir.

\section{Marginal Distributions, the Wigner Distribution and the Inverse Radon Transform}
\label{sec:radon}

In the analysis of quadrature statistics, we consider the set of marginal distributions $e^{-\phi(x^\alpha)}$, associated with the general quadratures defined in Eq.~\eqref{eq:generalquad}, where $\alpha$ varies over the range $[0,\pi]$.  We also employ the quasi-probability distribution of the light, known as the Wigner distribution~\cite{gerry2005introductory,Ariano2,gardiner2004}.  We use these distributions to study the light in the reservoir coupled to the system. 

Taking the density matrix of the reservoir, ${\rho}_{\text{res}}$,  the Wigner function in the Schr\"odinger representation is
\vspace*{-5mm}
\begin{center}
\begin{equation}\label{11e}
{W}_{\rho}\left(\vartheta,{\vartheta}^{*}\right)=\frac{1}{{\pi}^{2}}\int {d}^{2} \delta\:{e}^{-\delta{\vartheta}^{*}+ {\delta}^{*}\vartheta}{\text{Tr}}_{\text{res}}\left({\rho}_{\text{res}}{e}^{\left(\delta{b}^{\dag}-{\delta}^{*}b\right)}\right)\,.
\end{equation}
\end{center}
The variable $\vartheta$ is identified with the quadratures through the relations
\begin{eqnarray}
X &=& \text{Re}\,(\vartheta) \\
Y &=& \text{Im}\,(\vartheta)\,.
\end{eqnarray}
As direct evaluation of this distribution is difficult we will use the fact that ${W}_{\rho}$ is related to the marginals $e^{-\phi(x^\alpha)}$ by the Radon transform~\cite{Revzen}, $\mathcal{R}$, via
\vspace*{-5mm}
\begin{center}
\begin{align}\label{12e}
&e^{-\phi(x^\alpha)}=\mathcal{R}[{W}_{\rho}]\\
&=\int^{+\infty}_{-\infty}{W}_{\rho}\left(X\cos\alpha\!-\! Y\sin\alpha,Y\cos\alpha\!+\! X\sin\alpha\right)dY\nonumber
\end{align}
\end{center}
and we employ its numerical inverse to the marginals $$e^{-\phi(x^\alpha)}$$ to obtain ${W}_{\rho}$.  This technique is frequently used in quantum tomography~\cite{Ariano2}. The Wigner distribution allows classification of the light into quantum and classical components based on the sign of the distribution: negative regions are indicative of quantum effects such as entanglement.

\section{Full diagonalization of the $s$-biased generalised master equation for the micromaser}
\label{sec:DIAG}
To determine the LD function of the micromaser numerially,  we numerically diagonalized the generalized master operator ${\mathcal{W}}_{s}$ in matrix form, as in Ref.~\cite{Genway1}. It is necessary to truncate the  basis of the system. The basis of number states $\ket{n}$ is suitable and, for the ${N}_{ex}$ value of 100 studied in this work, we restrict the maximum photon number to $n=150$. The form of ${\mathcal{W}}_{s}$ introduces coherences between different eigenstates of the cavity and so we must ensure our basis allows these coherences to be preserved.  However, since in matrix form the representation of the superoperator is an $n^2 \times n^2$ matrix, we investigated whether further truncation was possible.  In practice it is possible to truncate the basis in such a manner that only coherences between number states with occupation number differing by $m$ (with $m<n$) are preserved. The value of $m$ may be tested numerically to ensure the results are not sensitive to this truncation.  In the results presented in this work, the values of $n=150$ and $m=15$ were found to be sufficient for the diagonalization process. The largest real eigenvalue extracted from this matrix is identified as $\theta\left(s\right)$.  The matrices were diagonalized using an Arnoldi iteration scheme~\cite{Arpack}. The activity was calculated by taking numerical derivatives of the resulting LD function.

\end{document}